# Novel Algorithm for Computing All-Pairs Homogeneity-Arc Binary-State Undirected Network Reliability

Wei-Chang Yeh
Department of Industrial Engineering and Engineering Management
National Tsing Hua University
P.O. Box 24-60, Hsinchu, Taiwan 300, R.O.C.
yeh@ieee.org

*Abstract* — Among various real-life emerging applications, wireless sensor networks, Internet of Things, smart grids, social networks, communication networks, transportation networks, and computer grid systems, etc., the binary-state network is the fundamental network structure and model with either working or failed binary components. The network reliability is an effective index for assessing the network function and performance. Hence, the network reliability between two specific nodes has been widely adopted and more efficient network reliability algorithm is always needed. To have complete information for a better decision, all-pairs network reliability thus arises correspondingly. In this study, a new algorithm called the all-pairs BAT is proposed by revising the binary-addition-tree algorithm (BAT) and the layered-search algorithm (LSA). From both the theoretical analysis and the practical experiments conducted on 20 benchmark problems, the proposed all-pairs BAT is more efficient than these algorithms by trying all combinations of any pairs of nodes.

*Keywords*:   Binary-state Network; Network Reliability; All-Pairs; Binary-Addition-Tree Algorithm (BAT), Layered-search algorithm (LSA); Undirected Arc; Homogenous Arc

## 1. INTRODUCTION

Among various real-life emerging applications, the binary-state network is the fundamental structure and each component is either working or failed. For example, the network transmission problems involving (signals) communication [1], (liquids or gases) distribution [2,3], (traffic) transportation [4], (topology) transformation [5, 6, 7], (power) transmission [8]; grid/cloud computing [9]; data mining [10]; Internet of Things [11, 12]; and network resilience problems [13, 14] all can be modeled as binary-state networks. All components (arcs and/or nodes) in binary-state networks

for arXiv

composed of binary states, i.e., working or failed [15, 16].

Researches and applications of the binary-state network, thus, have been increasingly in development, strategy, implementation, managing, and control for all the above-mentioned systems recently. Variations in applications of the modern network are increasing in number gradually, and different related applications are accordingly wider and broader [16, 17, 18]. The network reliability of a pair of specific nodes is the probability that these two nodes are connected by any path. The reliability of network is an operative and general technique to calculate (how likely is the network to be still effective) even if it is a NP-hard to calculate the reliability of any binary-state network [16, 17, 18].

However, the traditional network reliability problem mainly focuses only on the connectivity between a pair of specific nodes. For decision makers, the reliability of only a pair of nodes is not enough to validate the network performance and roles of components to make a correct decision. For example, a decision maker is always needed to find which pair of locations is the most reliable to build a train track in the transportation system [4]; which two bases have the best reliable connection after considering network resiliency in telecom system [13]; which pairs has the less reliable communication in the wireless sensor network [14]; which two grids need to enhance their connection to have a better reliable smart grid [9]; which points have the weakest signal transmission in the internet of things [11, 12], etc.

In the proposed all-pairs homogeneity-arc binary-state undirected network reliability problem, the desired output is a symmetry matrix of which the $i$th row and the $j$th column represent the reliability between nodes $i$ and $j$ for all nodes $i$ and $j$.

The binary-state network reliability can already be estimated by using many tools and approaches [19-25]. It is trivial that the all-pairs homogeneity-arc binary-state undirected network reliability problem can be solved by applying any algorithm that calculates the binary-state undirected network reliability between all pair of nodes. However, the running time of the above concept is $O(m(m-1)\rho)$, where $m$ is the number of arcs and $\rho$ is the time complexity of a traditional binary-state undirected 3network reliability algorithm, e.g., $\rho = O(m2^{2m})$ for the BAT proposed in [25-29]. From the application purpose, a more efficient algorithm is always necessary to calculate the exact all-pairs binary-state network reliability.

The traditional directed-arc binary-addition-tree algorithm (BAT) proposed by Yeh in [25] has emerged recently and became a new search method in finding all possible solutions or vectors. BAT is simply based on the binary addition to find all solutions and is more efficient compared to the depth-first search and breadth-first search methods. BAT has also been applied in different applications, i.e., the network reliability problems [25, 26, 27], the spread of wildfire [28], the propagation of computer virus [29], and rework problems [30].

Furthermore, for the network reliability problems, BAT has been implemented to solve different type of networks, e.g., binary-state networks of which each component is either working or failed [25, 27], multi-state networks of which each component has different capacities or states with different probabilities [29], flow networks which satisfied the flow conservation law, and/or information networks which dissatisfied the flow conservation law [26].

In calculating the binary-state network reliability, BAT can calculate and sum up the probabilities of all connected vectors to have the final reliability directly. Hence, BAT has no need to find out neither all minimal paths nor minimal cuts which both are NP-Hard problems, and neither to use Inclusion-Exclusion Technique (IET) [31] nor Sum-of-disjoint Product (SDP) [32] which both are also NP-Hard problems to calculate the reliability in terms of minimal paths or minimal cuts [17, 18]. Thus, from the complete experiments, the BAT is more efficient than these indirect algorithms.

Because its efficiency, simplicity, and flexibility of the BAT [25], the goal of this study is to revise the traditional directed-arc BAT to be an all-pairs undirected-arc BAT for the proposed all-pairs network reliability problem.

The rest of this study is organized as follows. Section 2 contains acronyms, notations, nomenclatures, and assumptions. The preliminaries, in particular, the concepts, pseudocode, and examples of the traditional BAT and path-based layered-search algorithm (PLSA) are discussed in Section 3. The major contributions, including the utility of the undirected arcs rather than the directed arcs, the novel connected-group layered-search algorithm (CG-LSA) in verifying the all-pairs connectivity efficiently, and the pre-calculation of the probability to improve the running time of the all-pairs undirected BAT are proposed in Section 4. The primary difference between the proposed algorithm and the traditional BAT, the pseudocode, an example, and the time complexity of the proposed BAT are discussed in Section 5. Furthermore, the performance of the proposed BAT was demonstrated by testing on 20 benchmark networks. We conclude this study in Section 6.

## 2. ACRONYMS, NOTATIONS, NOMENCLATURES, AND ASSUMPTIONS

Relevant acronyms, notations, assumptions, and nomenclatures are presented in this section.

### 2.1 Acronyms

- BAT: binary-addition-tree algorithm [25]
- LSA: layered-search algorithm [32]
- CG-LSA: connected-group LSA
- BDA: binary-state directed arc
- BUA: binary-state undirected arc

### 2.2 Notations

- $/\bullet/$: number of elements in set $\bullet$ or the number of coordinates with value 1 in vector $\bullet$
- $n$: numbers of nodes
- $m$: numbers of arcs
- $V$: complete node set $V = \{1, 2, \ldots, n\}$
- $E$: complete arc set $E = \{a_1, a_2, \ldots, a_m\}$
- $a_k$: $k^{\text{th}}$ undirected arc in $E$
- $a_{i,j}$: $a_{i,j} \in E$ and $a_{i,j} = a_{j,i} = a_k$ for one $k$ and $i, j \in V$
- $e_{i,j}$: directed arc $e_{i,j} \in E$ between nodes $i$ and $j$ note that $e_{i,j} \neq e_{j,i}$
- $\Pr(\bullet)$: probability of the occurrence of event $\bullet$

**D**: state distributions of homogeneous undirect arcs list $P(a)$ for each arc $a$

$G(V, E)$: A graph with $V$ and $E$, e.g., the bridge network in Fig. 1 shows a graph with $V = \{1, 2, 3, 4\}$ and $E = \{a_1 = e_{1,2}, a_2 = e_{1,3}, a_3 = e_{2,3}, a_4 = e_{2,4}, a_5 = e_{3,4}\}$.

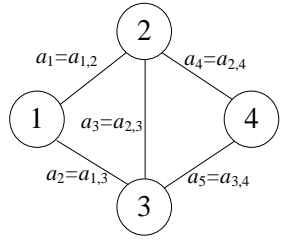

**Figure 1.** The bridge network

$G(V, E, \mathbf{D})$: A homogeneity-arc binary-state undirected network with graph $G(V, E)$ and state distributions $\mathbf{D}$, e.g., $G(V, E)$ in Fig. 1 with $\mathbf{D}$ presented in Table 1 is a $G(V, E, \mathbf{D})$.

**Table 1.** Arc state distributions in Fig. 1

| $e$ | $a_1 = a_{1,2}$ | $a_2 = a_{1,3}$ | $a_3 = a_{2,3}$ | $a_4 = a_{2,4}$ | $a_5 = a_{3,4}$ |
|---|---|---|---|---|---|
| Pr($e$) | 0.9 | 0.8 | 0.7 | 0.6 | 0.5 |

$X$: binary-state vector and the value of its $i^{th}$ coordinate is the state of arc $a_k \in E$ for $k = 1, 2, \ldots, m$

$X(a_k)$: state (value) of the $k^{th}$ coordinate $a_k \in E$ in the binary-state vector $X$ for $k = 1, 2, \ldots, m$

Pr($X(a_i)$): occurrence probability of $a$ in vector $X$

Pr($X$): $\Pr(X) = \prod_{k=1}^{m} \Pr(X(a_k))$

$G(X)$: subgraph $G(X) = G(V, \{a \in E \mid \text{for all arc } a \text{ with } X(a) = 1\})$

P($i$): $P(i) = p^i \times (1-p)^{(m-i)}$

$\Omega_{s,t}(G)$: $\Omega_{s,t}(G) = \{X \mid X \text{ is a } (s, t)\text{-connected vector in } G(V, E, \mathbf{D})\}$

$R_{s,t}(G)$: $R_{s,t}(G) = \Pr(\Omega_{s,t}(G))$ is reliability between nodes $a$ to $b$ of $G(V, E, D)$

### 2.3 Nomenclatures

Paired reliability: The success probability that there is one direct path between two specific nodes

All-pairs reliability: All paired reliabilities

($s$, $t$)-connected vector: A vector $X$ is connected if there is at least one directed path from nodes s to t in G(X).

Homogenous Arc: An arc between nodes $i$ and $j$ is homogenous if its reliability from node $a$ to $b$ is identical to that from nodes $j$ to $i$, i.e., $e_{i,j} \in E$ if and only if $e_{j,i} \in E$ and also $\Pr(e_{i,j}) = \Pr(e_{j,i})$.

**2.4 Assumptions**

1. Each node is completely reliable in $V$.

2. Each arc is undirected, homogenous, and either working or failed in $E$.

3. $G(V, E)$ is connected without parallel arcs or loops.

4. $\Pr(a)$ is statistically independent according to a given distribution in $D$ for each $a$ in $E$.

**3. REVIEW OF BAT AND PLSA**

The traditional (directed-arc) BAT is revised in the proposed algorithm to find all possible undirected vectors and the PLSA is modified to verify the connectivity between any pairs of nodes. Hence, the traditional BAT and the PLSA are presented in this section.

**3.1 BAT**

The original BAT proposed by Yeh [25] is based on the binary addition for the binary-state directed network reliability problems. BAT and its variants have been compared with the breadth-search-first algorithm (BFS), e.g., the universal generating function methodology (UGFM) [16, 26] which is the best-known algorithm in the multi-state information networks without satisfying the flow conservation law; the depth-search-first algorithms (DFS), e.g., the QIE [31] which outperforms the best-known algorithm in multi-state flow networks reliability problems: the recursive BFS-based SDP (RSDP) [20, 32, 33]; the binary-decision-diagram (BBD) which is the best-known algorithm in binary-state network reliability problems [21, 27].

From performances in experiments, BAT algorithms outperform BFS [16, 26], DFS [31], UGFM [16, 26], QIE [31], RSDP [32], and BBD [21] in both running time without having an overflow for arXiv problem in computer memory. Furthermore, BAT is simple to understand, easy to code, and convenient to make-to-fit [25-30].

The original BAT is implemented in this paper to find all possible combinations of each binary-state directed arc (BDA) vector [25]. In any BDA vector, each coordinate value is either 0 or 1 to represent that the related arc is failed or functioning, respectively. In BAT, the initial BDA vector is vector zero, i.e., each coordinate value is zero. In the original (backward) BAT [25], the procedure to find all BDA vectors is by adding one to the last coordinate to change the coordinate value from the last to the first iteratively, in a backward way, based on the binary addition method [25]:

1. If the value of the current coordinate is 0, it is changed to 1 and a new BDA vector is found. For example, (0, 0, 1) is updated to (0, 1, 0), i.e., 001 is updated to 010 by adding 1 to the last coordinate, where (0, 0, 1) is the current found BDA vector and (0, 1, 0) is the next found BDA vector.

2. If the value of the current coordinate, say $a_k \in E$, is 1, then it is changed to 0 and move to the next coordinate, i.e., $a_{(k-1)}$, to repeat the above two steps. For example, (1, 0, 1) is updated to (1, 1, 0), i.e., 101 is updated to 110 by adding 1 to the last coordinate.

**3.2 BAT Pseudo-code and Time Complexity**

The pseudo-code of the original (backward) BAT for finding all BDA vectors between nodes $s$ and $t$ is provided as follows [25]:

**Algorithm: BAT**

**Input:** A directed graph $G(V, E)$ with the source node $s$ and the sink node $t$.

**Output:** All BDA vectors without duplications.

**STEP B0.** Remove $e_{i,s}$ and $e_{t,j}$ for all $e_{i,s} \in E$ and $e_{t,j} \in E$, and let $m^* = |\ E - \{\ e_{i,s},\ e_{t,j}$ for all $e_{i,s} \in E$ and $e_{t,j} \in E\ \}|$.

**STEP B1.** Assign SUM = 0, $X$ be a vector zero, and $k = m^*$.

**STEP B2.** If $X(a_k) = 1$, assign $X(a_k)$ and SUM to be 0 and SUM − 1, respectively, and go to STEP B5.

**STEP B3.** Assign $X(a_k)$ and SUM to be 1 and SUM + 1, respectively.

**STEP B4.** Let $k = k − 1$ and return to STEP B2 if $k > 1$.

**STEP B5.** If SUM $= m^*$, halt; otherwise, let $k = m^*$ and return to STEP B2.

STEP B0 replaces each undirected arc with two directed arc and removes these arcs which are impossible to go through. STEP B1 initializes the BDA vector $X$, the number of coordinates with one, and the current index $k$ to be a vector zero, value zero, and $m^*$ (the last coordinate), respectively. These two procedures in generating a new BDA vector listed before are given from STEPs B2 and B3-B4, respectively. STEP B5 is the stopping criteria.

Note that there is no need to save the current $X$ if we used it to generate a new BDA vector and determine its corresponding values, e.g., probability, cost, time, or any predefined function [25].

The number of all found $X$ is $2^{2m}$ after changing each undirected arc into two directed arcs with opposite directions without considering STEP B0. The total time complexity is $O(n2^{2m+1})$, where $O(n)$ is the running time for the PLSA [25], $O(2^{2m+1})$ is the time complexity of the traditional directed-arc BAT, and the details of the time complexity of BAT is discussed in Section 5.2.

**3.3 BAT Example**

The bridge network depicted in Fig. 1 is used to demonstrate the traditional directed-arc BAT. As listed in STEP B0, the only undirected arc $a_{2,3}$ is replaced with directed arcs $e_{2,3}$ and $e_{3,2}$ in Fig. 2 after removing the directed arcs $e_{i,0}$ and $e_{4,j}$ for nodes $i$ and $j$.

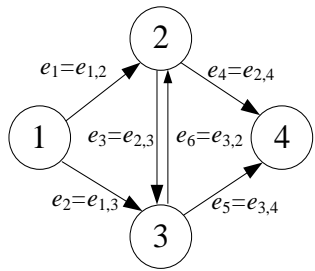

for arXiv

**Figure 2**. $a_{2,3}$ is replaced with $e_{2,3}$ and $e_{3,2}$ in Fig. 1.

After that, $m^* = 6$, i.e., each vector $X$ is a 6-tube vector. Let $X = (0, 0, 0, 0, 0, 0)$ and SUM = 0 based on STEP B1. Using the binary addition to the last coordinate in $X$, i.e., in backward way, by treating $X$ as a binary digit, i.e., 000000, after adding one, we have the new $X = (0, 0, 0, 0, 0, 1)$ from STEP B2. In the same way, we have all 64 vectors with no duplications as shown in Table 2.

**Table 2.** All 64 vectors obtained from the traditional directed-arc BAT.

| $i$ | $X_i$ | $i$ | $X_i$ | $i$ | $X_i$ | $i$ | $X_i$ |
|---|---|---|---|---|---|---|---|
| 1 | (0, 0, 0, 0, 0, 0) | 17 | (0, 1, 0, 0, 0, 0) | 33 | (1, 0, 0, 0, 0, 0) | 49 | (1, 1, 0, 0, 0, 0) |
| 2 | (0, 0, 0, 0, 0, 1) | 18 | (0, 1, 0, 0, 0, 1) | 34 | (1, 0, 0, 0, 0, 1) | 50 | (1, 1, 0, 0, 0, 1) |
| 3 | (0, 0, 0, 0, 1, 0) | 19 | (0, 1, 0, 0, 1, 0) | 35 | (1, 0, 0, 0, 1, 0) | 51 | (1, 1, 0, 0, 1, 0) |
| 4 | (0, 0, 0, 0, 1, 1) | 20 | (0, 1, 0, 0, 1, 1) | 36 | (1, 0, 0, 0, 1, 1) | 52 | (1, 1, 0, 0, 1, 1) |
| 5 | (0, 0, 0, 1, 0, 0) | 21 | (0, 1, 0, 1, 0, 0) | 37 | (1, 0, 0, 1, 0, 0) | 53 | (1, 1, 0, 1, 0, 0) |
| 6 | (0, 0, 0, 1, 0, 1) | 22 | (0, 1, 0, 1, 0, 1) | 38 | (1, 0, 0, 1, 0, 1) | 54 | (1, 1, 0, 1, 0, 1) |
| 7 | (0, 0, 0, 1, 1, 0) | 23 | (0, 1, 0, 1, 1, 0) | 39 | (1, 0, 0, 1, 1, 0) | 55 | (1, 1, 0, 1, 1, 0) |
| 8 | (0, 0, 0, 1, 1, 1) | 24 | (0, 1, 0, 1, 1, 1) | 40 | (1, 0, 0, 1, 1, 1) | 56 | (1, 1, 0, 1, 1, 1) |
| 9 | (0, 0, 1, 0, 0, 0) | 25 | (0, 1, 1, 0, 0, 0) | 41 | (1, 0, 1, 0, 0, 0) | 57 | (1, 1, 1, 0, 0, 0) |
| 10 | (0, 0, 1, 0, 0, 1) | 26 | (0, 1, 1, 0, 0, 1) | 42 | (1, 0, 1, 0, 0, 1) | 58 | (1, 1, 1, 0, 0, 1) |
| 11 | (0, 0, 1, 0, 1, 0) | 27 | (0, 1, 1, 0, 1, 0) | 43 | (1, 0, 1, 0, 1, 0) | 59 | (1, 1, 1, 0, 1, 0) |
| 12 | (0, 0, 1, 0, 1, 1) | 28 | (0, 1, 1, 0, 1, 1) | 44 | (1, 0, 1, 0, 1, 1) | 60 | (1, 1, 1, 0, 1, 1) |
| 13 | (0, 0, 1, 1, 0, 0) | 29 | (0, 1, 1, 1, 0, 0) | 45 | (1, 0, 1, 1, 0, 0) | 61 | (1, 1, 1, 1, 0, 0) |
| 14 | (0, 0, 1, 1, 0, 1) | 30 | (0, 1, 1, 1, 0, 1) | 46 | (1, 0, 1, 1, 0, 1) | 62 | (1, 1, 1, 1, 0, 1) |
| 15 | (0, 0, 1, 1, 1, 0) | 31 | (0, 1, 1, 1, 1, 0) | 47 | (1, 0, 1, 1, 1, 0) | 63 | (1, 1, 1, 1, 1, 0) |
| 16 | (0, 0, 1, 1, 1, 1) | 32 | (0, 1, 1, 1, 1, 1) | 48 | (1, 0, 1, 1, 1, 1) | 64 | (1, 1, 1, 1, 1, 1) |

**3.4 PLSA**

In the original BAT for the binary-state directed network reliability problems, the path-based layered-search algorithm (PLSA) was implemented in STEP B2 in Section 3.1 to verify whether the new obtained BDA vector $X$ is a ($s$, $t$)-connected vector, i.e., there is a directed path from nodes $s$ to $t$. If such $X$ is a ($s$, $t$)-connected BDA vector, Pr($X$) is calculated; otherwise, go to B5.

The PLSA is revised from the layered-search algorithm (LSA) proposed in [34] to search for all $d$-MPs in acyclic networks. The PLSA pseudocode is delivered as follows.

**Algorithm: PLSA**

**Input:** A BDA vector obtained $X$ in STEP B2 of BAT

for arXiv

**Output:** Whether $X$ is $(s, t)$-connected from nodes $s$ to $t$ in $G(V, E)$, where $s = 1$ and $t = n$.

**STEP P0.** Let $V^* = L_1 = \{1\}$ and $i = 2$.

**STEP P1.** Let $L_i = \{ v \notin V^* \mid$ for all $e_{\alpha,v} \in E$ and $\alpha \in V^*$, i.e., $X(a_i)=1$ $a_i = e_{\alpha,v}\}$.

**STEP P2.** If $n \in L_i$, there is a directed $(1, n)$-path in $G(X)$, i.e., $X$ is $(s, t)$-connected, and stop.

**STEP P3.** If $L_i = \emptyset$, there is a cut between nodes $s$ and $t$ in $G(X)$ and stop.

**STEP P4.** Let $V^* = V^* \cup L_i$, $i = i + 1$, and go to STEP P1.

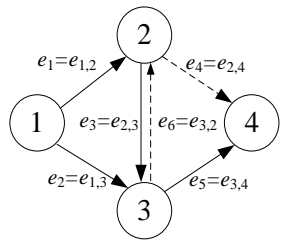

**Figure 3.** $G(X)$ based on Fig. 2, where $X = (1, 1, 1, 0, 1, 0)$.

For example, let $X = (1, 1, 1, 0, 1, 0)$ in Fig. 2 and $G(X)$ is shown in Fig. 3. The process of verifying the connectivity between nodes 1 and 4 in $G(X)$, i.e., to determine whether a BDA vector $X$ is a $(1, 4)$-connected, based on PLSA is shown in Table 3.

**Table 3.** Verify the connectivity between nodes 1 and 4 by using PLSA for Fig. 1.

| $i$ | $L_i$ | $V^*$ |
|---|---|---|
| 0 | {1} | {1} |
| 1 | {2, 3} | {1, 2, 3} |
| 2 | {4, 5} | {1, 2, 3, 4, 5} |

## 4. INNOVATION PARTS IN PROPOSED ALL-PAIRS BAT

The innovation parts in proposed all-pairs BAT are discussed in this section including the reason why there is no need to change an undirected arc into two directed arcs [25]; the novel concept to verify the connectivity of all pairs of nodes by just using one complete vector set obtained from the BAT without generating all vector sets for each pair of nodes to save running time; another new

for arXiv

concept to calculate the probability of possible vectors based on the number of working homogenous arcs before finding all vectors to reduce the computer burden.

**4.1 Undirected arcs**

The number of BDA vectors is the major part in the BAT time complexity. In the traditional BAT, any undirected arc is needed to change to two directed arcs with two opposite directions. Hence, the number of coordinates is $2m$ and the time complexity is $O(2^{2m})$ for the traditional BAT to have all BDA vectors [25], where the power of two denotes that each directed arc has two states. Hence, if we can reduce the number of coordinates, the time complexity can be also reduced. To achieve the above goal, the following property is proposed and proved.

**Property 1.** $R_{s,t}(G) = R_{s,t}(G_k)$, where $G = G_k$ except the undirected arc $a_k = a_{i,j}$ in $G$ are changed to two direct arcs with two opposite directions: $e_{x,y}$ and $e_{y,x}$ with arc labels $k$ and the $(m+1)$ in $G_k$.

**Proof.** Let vector subset $\Xi$ and $\Xi_k$ be the sets of all $m$-tube and $(m+1)$-tube vectors such that

$$\Xi = \{ X \mid X(a_i) = x_i \text{ for } i \neq k \} \tag{1}$$

and

$$\Xi_k = \{ X \mid X(a_i) = x_i \text{ for } i \neq k, (m+1) \}. \tag{2}$$

If any vector in $\Xi$ is disconnected, then any vector in $\Xi_k$ is also disconnected and both $\Pr(\Xi)$ and $\Pr(\Xi_k)$ are not counted in $R_{s,t}(G)$ and $R_{s,t}(G_k)$, respectively. Conversely, if any vector in $\Xi$ is connected and no matter the state of $a_k$, i.e., there is a path via $e_{x,y}$ and another path via $e_{y,x}$ from nodes $s$ to $t$, both $\Pr(\Xi)$ and $\Pr(\Xi_k)$ are counted in $R_{s,t}(G)$ and $R_{s,t}(G_k)$ and

$$\Pr(\Xi) = \Pr(\{X \in \Xi \mid X(a_k)=1\}) + \Pr(\{X \in \Xi \mid X(a_k)=0\}) = \prod_{\substack{i=1 \\ i \neq k}}^{m} \Pr(a_i) \tag{3}$$

$$\Pr(\Xi_k) = \Pr(\{K \in \Xi_k \mid K(e_{x,y})=K(e_{y,x})=1\})$$
$$+ \Pr(\{K \in \Xi_k \mid K(e_{x,y})=K(e_{y,x})=0\})$$

for arXiv

$$+ \Pr(\{K \in \Xi_k \mid K(e_{x,y})=1, K(e_{y,x})=0\})$$

$$+ \Pr(\{K \in \Xi_k \mid K(e_{x,y})=0, K(e_{y,x})=1\})$$

$$= \prod_{\substack{i=1 \\ i \neq k}}^{m} \Pr(a_i). \tag{4}$$

Consider that any vector $X \in \Xi$ is connected by at least one path from nodes $s$ to $t$ via $e_{x,y}$ in $G(X_k)$. It is trivial that $\Pr(\Xi)$ is included in $R_{s,t}(G)$ and $\Pr(\Xi_k)$ with $X^*(e_{x,y}) = 1$ is included in $R_{s,t}(G_k)$ for all $K \in \Xi_k$, respectively, and

$$\Pr(\Xi) = \Pr(\{X \mid X(a_k)=1\}) = p \prod_{\substack{i=1 \\ i \neq k}}^{m} \Pr(a_i) \tag{5}$$

$$\Pr(\Xi_k) = \Pr(\{K \mid K(e_{x,y})=1, K(e_{y,x})=0\}) + \Pr(\{K \mid K(e_{x,y})=K(e_{y,x})=1\})$$

$$= pq \prod_{\substack{i=1 \\ i \neq k}}^{m} \Pr(a_i) + pp \prod_{\substack{i=1 \\ i \neq k}}^{m} \Pr(a_i)$$

$$= p \prod_{\substack{i=1 \\ i \neq k}}^{m} \Pr(a_i). \tag{6}$$

Similarly, $\Pr(\Xi) = \Pr(\Xi_k)$, if any vector $X \in \Xi$ is connected and at least one path from nodes $s$ to $t$ via $e_{y,x}$ in $G(X_k)$. From the above, this property is correct.

From the above discussion in Property 1, the number of coordinates can be reduced from $2m$ to $m$ in the proposed all-pairs BAT without changing the final reliability of any pair of nodes.

**4.2 Connected-Group Layered-Search Algorithm (CG-LSA)**

The original PLSA can only verify the connectivity of a pair of nodes. In the proposed all-pairs BAT, we need to verify the connectivity of all pairs. Hence, the PLSA which is path-based is modified to a new algorithm called the connected-group layered-search algorithm (CG-LSA).

The pseudocode of the proposed CG-LSA in verifying the connectivity of all pairs of nodes is listed below:
for arXiv

**ALGORITHM:** CG-LSA

**Input:**  A BUA $X$ vector obtained in the proposed all-pairs BAT

**Output:**  Which nodes are connected in $G(X)$.

**STEP S0.** Let $\tau = i = 0$ and $f_v = 0$ for all $v \in V$.

**STEP S1.** Find the node, say $k$, with $f_k = 0$, let $f_k = 1$, $\tau = \tau + 1$, and $L = V_\tau = \{k\}$. If no such node, halt.

**STEP S2.** Let $i = i + 1$, $k$ be the first node in $L$, $L_{\tau,i} = \{ v \in V \mid $ for all $e_{k,v} \in E$ and $f_v = 0$, i.e., $X(a_j) = 1$ and $a_j = e_{k,v}\}$, $V_\tau = V_\tau \cup \{k\}$, and $L = (L - \{k\}) \cup L_{\tau,i}$.

**STEP S3.** If $L \neq \varnothing$, go to STEP S2. Otherwise, go to STEP S1.

Let $L_{\tau,i}$ be the $i$th layer, where $\tau$ denotes the $\tau$th connected group. $L_{\tau,i}$ is a special node subset and each node in such layer is connected to all nodes in $L_{\tau,k}$ for $k \leq i$. $L_{\tau,i}$ in STEP S2 be the core of the CG-LSA, we select another new node based on $L_{\tau,i}$ to form a new iteration after adding all nodes in $L_{\tau,i}$ to $L$ as previously shown in STEP S2. If $L_{\tau,i} = \varnothing$ and no more new nodes are left in $L$, we let $\tau = \tau + 1$ and select the node, say $k$, with smallest label and $f(k) = 1$, i.e., $f(j) = 0$ for all $j < k$, as provided in STEP S1.

Each node is selected to be node $k$ once only either in STEP S1 or STEP S2. Hence, the time complexity of the proposed CG-LSA is only $O(3n)$ including the update of $f$ and $\tau$. The well-known Floyd-Warshall algorithm [35, 36] found all-pairs shortest paths can also be used to verify the connectivity between any two nodes with time complexity $O(n^3)$. Also, the PLSA can be implemented by verifying all combinations of any two nodes. The number of different combinations is $P_2^n = n(n-1)/2$ and each time has time complexity $O(n)$, i.e., the total time complexity is also $O(n^3)$. Hence, the proposed CG-LSA is very efficient from the above comparisons.

for arXiv

For example, in Fig. 4 in which all dashed arcs are failed, the proposed SP-LST is implemented to find the connected groups and the whole procedure is explained in Table 4, where F = ($f_1, f_2, f_3, f_4, f_5, f_6, f_7, f_8, f_9, f_{10}, f_{11}, f_{12}, f_{13}, f_{14}, f_{15}, f_{16}$).

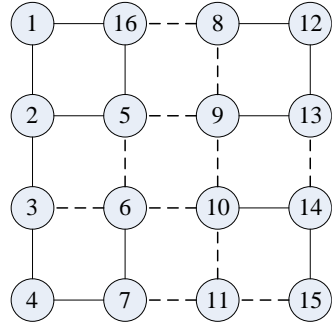

**Figure 4.** The example network to explain the proposed CG-LSA where all dashed arcs are failed.

To be easily recognized and understood, in Table 4, the number in bold and in a box denotes the related node $k$ selected in the current iteration and for the next iteration with a new $\tau$, respectively. For example, F = (1, 1, 1, 1, 1, **1**, 1, 0̄, 0, 0, 0, 0, 0, 0, 0, 1) for $i = 8$ denotes that node $k = 6$ is selected for the current iteration, i.e., the 8$^{th}$ iteration, and node 8 is going to be selected for the next iteration with $\tau = 2$ because $L = \varnothing$ in the end of the current iteration.

**Table 4.** The process from the proposed CG-LSA for Fig. 4.

| $i$ | $\tau$ | $k$ | $L_{\tau,i}$ | $L$ | F | $V_\tau$ |
|---|---|---|---|---|---|---|
| 0 | 0 |  | {1} | {1} | (1, 0, 0, 0, 0, 0, 0, 0, 0, 0, 0, 0, 0, 0, 0) |  |
| 1 | 1 | 1 | {2,16} | {2,6} | (**1**, 0, 0, 0, 0, 0, 0, 0, 0, 0, 0, 0, 0, 0, 0) | {1} |
| 2 | 1 | 2 | {3,5} | {16,3,5} | (1, **1**, 0, 0, 0, 0, 0, 0, 0, 0, 0, 0, 0, 0, 0) | {1, 2} |
| 3 | 1 | 16 | ∅ | {3,5} | (1, 1, 0, 0, 0, 0, 0, 0, 0, 0, 0, 0, 0, 0, **1**) | {1, 2, 16} |
| 4 | 1 | 3 | {4} | {5,4} | (1, 1, **1**, 0, 0, 0, 0, 0, 0, 0, 0, 0, 0, 0, 1) | {1, 2, 16, 3} |
| 5 | 1 | 5 | ∅ | {4} | (1, 1, 1, 0, **1**, 0, 0, 0, 0, 0, 0, 0, 0, 0, 1) | {1, 2, 16, 3, 5} |
| 6 | 1 | 4 | {7} | {7} | (1, 1, 1, **1**, 1, 0, 0, 0, 0, 0, 0, 0, 0, 0, 1) | {1, 2, 16, 3, 5, 4} |
| 7 | 1 | 7 | {6} | {6} | (1, 1, 1, 1, 1, 0, **1**, 0, 0, 0, 0, 0, 0, 0, 1) | {1, 2, 16, 3, 5, 4, 7} |
| 8 | 1 | 6 | ∅ | ∅ | (1, 1, 1, 1, 1, **1**, 1, 0̄, 0, 0, 0, 0, 0, 0, 1) | {1, 2, 16, 3, 5, 4, 7, 6} |
| 9 | 2 | 8 | {12} | 8 | (1, 1, 1, 1, 1, 1, 1, **1**, 0, 0, 0, 0, 0, 0, 1) | {8} |
| 10 | 2 | 12 | {13} | {13} | (1, 1, 1, 1, 1, 1, 1, 0, 0, 0, **1**, 0, 0, 0, 1) | {8, 12} |
| 11 | 2 | 13 | {9} | {9} | (1, 1, 1, 1, 1, 1, 1, 0, 0, 0, 1, **1**, 0, 0, 1) | {8, 12, 13} |
| 12 | 2 | 9 | ∅ | ∅ | (1, 1, 1, 1, 1, 1, 1, 1, **1**, 0̄, 0, 1, 1, 0, 0, 1) | {8, 12, 13, 9} |
| 13 | 3 | 10 | {14} | {14} | (1, 1, 1, 1, 1, 1, 1, 1, **1**, 0, 1, 1, 0, 0, 1) | {10} |
| 14 | 3 | 14 | {15} | {15} | (1, 1, 1, 1, 1, 1, 1, 1, 1, 0, 1, 1, **1**, 0, 1) | {10, 14} |
| 15 | 3 | 15 | ∅ | ∅ | (1, 1, 1, 1, 1, 1, 1, 1, 1, 0̄, 1, 1, 1, **1**, 1) | {10, 14, 15} |
| 16 | 4 | 11 | ∅ | ∅ | (1, 1, 1, 1, 1, 1, 1, 1, 1, **1**, 1, 1, 1, 1, 1) | {11} |

for arXiv

From Table 4, we have four connected groups: $V_1 = \{1, 2, 16, 3, 5, 4, 7, 6\}$, $V_2 = \{8, 12, 13, 9\}$, $V_3 = \{10, 14, 15\}$, and $V_4 = \{11\}$. All nodes in the same group are connected in $G(X)$, e.g., any pairs of nodes in $\{10, 14, 15\}$ are connected.

**4.3 Pre-calculations the Vector Probability**

Similar to all BAT reliability-related algorithms [25, 26, 27], after each new BUA vector is found and its connectivity is confirmed in the proposed all-pairs BAT, we need to calculate its reliability immediately before it is abandoned. To speed up the whole procedure, a new idea is proposed to calculate the reliability problems with $\Pr(a)$ are identical for all $a \in E$, e.g., $\Pr(a_i) = 0.9$ in Table 1.

Let

$$\Omega_{s,t}(G) = \{X \mid X \text{ is a } (s, t)\text{-connected vector in } G(V, E, \mathbf{D})\} \tag{7}$$

$$P(i) = P(|X|) = p^i \times (1-p)^{(m-i)}, \tag{8}$$

where $i = |X|$ is the number of coordinates with value 1 in vector $X$. The value of $P(i)$ can be calculated before the BAT for all $i = 0, 1, \ldots, m$. For example, let $p = 0.9$ and $q = 1 - p = 0.1$ be the success and failure probabilities of each arc in Fig. 1, respectively. We have $P(i)$ for $i = 0, 1, \ldots, m$ as shown in Table 5.

**Table 5.** The values of $P(i)$ in Fig .1 if all arcs are all homogenous arcs with $p = 0.9$.

| $i$ | $P(i)$ |
|---|---|
| 0 | $0.9^0 \times 0.1^5 = 0.00001$ |
| 1 | $0.9^1 \times 0.1^4 = 0.00009$ |
| 2 | $0.9^2 \times 0.1^3 = 0.00081$ |
| 3 | $0.9^3 \times 0.1^2 = 0.00729$ |
| 4 | $0.9^4 \times 0.1^1 = 0.06561$ |
| 5 | $0.9^5 \times 0.1^0 = 0.59049$ |

The occurrence probability of the connected BUA vector $X$, i.e., $\Pr(X)$, obtained in the BAT in STEP A2 is the product of the probabilities of all failed arcs and all functioning arcs in $X$, i.e.,

$$\Pr(X) = \prod_{k=1}^{m} \Pr(X(a_k)). \tag{9}$$

Note that

for arXiv

$$\sum_{\forall X} \Pr(X) = 1 \tag{10}$$

for all BUA vector $X$ obtained in BAT no matter $X$ is connected or disconnected.

The reliability of the connectivity of nodes $s$ and $t$ of $G(V, E, \mathbf{D})$, i.e., $R_{s,t}(G)$, is the summation of all $\Pr(X)$ for all connected BUA vector $X$:

$$R_{s,t}(G) = \Pr(\Omega_{s,t}(G)) = \sum_{X \in \Omega_{s,t}(G)} \Pr(X). \tag{11}$$

Hence, after substituting Eq. (9) for $\Pr(X)$ to Eq. (11), we have

$$R_{s,t}(G) = \sum_{X \in \Omega_{s,t}(G)} \prod_{k=1}^{m} \Pr(X(a_k)). \tag{12}$$

Because all $\Pr(a) = p$ for all arc $a$ in the homogenous networks, we have

$$\Pr(X) = \mathrm{P}(|X|). \tag{13}$$

and

$$R_{s,t}(G) = \sum_{X \in \Omega_{s,t}(G)} \mathrm{P}(|X|). \tag{14}$$

For example, we have $\Pr(X) = 0.9^{13} \times 0.1^{11}$ if $p = 0.9$ and $q = 0.1$ are the success and failure probabilities of each arc, respectively, and $X$ is the related state vector of Fig. 4.

For the reliability between nodes $s$ and $t$, i.e., $R_{s,t}(G)$, Eq. (13) and Eq. (14) take only $O(1)$ and $O(2^{|\mu|})$ if $\mathrm{P}(|X|)$ is pre-calculated; otherwise, $O(m)$ and $O(m2^{|\mu|})$ in Eq. (9) and Eq. (12), respectively, where $\mu = 2^{|m|}$ is the number of $(s, t)$-connected vectors in the worst case. Moreover, the time complexities are increased up to $O(n^2)$ and $O(n^2 \times m2^{|\mu|})$ for all-pairs network reliability problems for these networks without homogenous arcs. Thus, Eq. (12) is suitable for these networks without homogenous arcs and Eq. (14) is very useful in homogenous networks to reduce the running time.

## 5. PROPOSED BAT

The proposed all-pairs BAT for calculating all-pairs reliability is presented in Section 5 including the discussion of the differences between the proposed all-pairs BAT and the traditional

BAT in Section 5.1, the pseudocode and time complexity of the all-pairs BAT in Section 5.2, an example to explain the all-pairs BAT in Section 5.3, and a complete experiment on 20 benchmark problems to demonstrate the performance of the all-pairs BAT in Section 5.4.

**5.1 Differences Between the All-Pairs BAT and the Traditional BAT**

The way to generate all possible BUA vectors in the proposed all-pairs BAT is the exact same to that in the traditional BAT [25] which is discussed in Section 3. The value of each coordinate in a BUA vector is either 0 or 1 to denote the state of the related arc and each BUA vector can be treated as a binary code.

In the proposed all-pairs BAT, each new BUA vector is generated by adding 1 to the first coordinate, i.e., in a forward way, of the current BUA vector which is slightly different to the backward way in the traditional BAT proposed in [25]. Besides, there are three major differences between the proposed all-pairs BAT and the traditional BAT:

1. No need to change each undirected arc to two directed arcs with opposite directions. The correctness of the above statement is provided in Section 4.1.

2. Each pair of nodes are needed to verify its connectivity in the proposed all-pairs BAT not only one pair of nodes in the traditional BAT. Note that the way to achieve this goal is based on the CG-LSA proposed in Section 4.2 which can reduce the time complexity from $O(n^3)$ to at least $O(3n)$.

3. The way to calculate the probability of each connected BUA vector is based on Eq. (14) in the proposed all-pairs BAT but not based on Eq. (12) as in the traditional BAT. Eq. (14) is calculated only once before the proposed all-pairs BAT and can speed up the whole procedure in calculating all-pairs reliability from $O(n^2 m 2^{|m|})$ to $O(m 2^{|m|})$ as discussed in Section 4.3 if the verification of the connectivity between any pair of nodes is ignored.

**5.2 Pseudocode and Time Complexity of All-Pairs BAT**

The overall pseudocode of the proposed BAT is based on the traditional BAT proposed in [25] without considering the directions of arcs as discussed in Section 4.1, the CG-LSA discussed in Section 4.2 in verifying the connectivity between any pair of nodes in the BUA vectors, and the pre-calculation of $P_i$ for $i = 0, 1, …, m$ discussed in Section 4.3 as follows.

**Algorithm:** All-Pairs BAT

**Input:** A homogeneity-arc binary-state undirected network $G(V, E, \mathbf{D})$

**Output:** The reliability $R_{s,t}(G)$ for all nodes $s \neq t$ in $V$

**STEP 0.** Let SUM = 0, $X = \mathbf{0}$ (vector zero), $k = 1$, and calculate $P(i)$ for $i = 0, 1, …, m$ if $\Pr(a) = \Pr(b)$ for all arcs $a \neq b$ in $E$.

**STEP 1.** If $X(a_k) = 1$, let $X(a_k) = 0$ and SUM = SUM − 1, and go to STEP 4.

**STEP 2.** Let $X(a_k) = 0$ and SUM = SUM + 1.

**STEP 3.** If $k < m$, let $k = k + 1$ and go to STEP 1.

**STEP 4.** Find $\Omega(X) = \{ (s, t) \mid$ if $X$ is a $(s, t)$-connected for all nodes $s \neq t$ in $V \}$ based on the proposed CG-LSA proposed in Section 4.3.

**STEP 5.** If $\Pr(a) = \Pr(b)$ for all arcs $a \neq b$ in $E$, let $R(s, t) = R(s, t) + P(|X|)$ for all $(s, t) \in \Omega(X)$ and go to STEP 7.

**STEP 6.** Let $R(s, t) = R(s, t) + \prod_{k=1}^{m} \Pr(X(a_k))$ for all $(s, t) \in \Omega(X)$.

**STEP 7.** If SUM = $m$, halt and $R_{s,t}(G) = R(s, t)$ for all nodes $s, t \in V$. Otherwise, go to STEP 1.

STEP 0 initializes all values, e.g., SUM, $X$, and $k$, and calculates $P(i)$ for all $i = 0, 1, …, m$ if $G(V, E, \mathbf{D})$ is homogeneous. STEPs 1 – 3 is based on the BAT to find each BUA vector. STEP 4 verifies the connectivity of each pair of nodes based on the proposed CG-LSA. STEPs 5 and 6 add the probability of $P(|X|)$ or $\Pr(X)$ to $R(s, t)$, respectively, if $X$ is a $(s, t)$-connected for each pair of nodes $s$ and $t$.

for arXiv

Note that no two BUA vectors have the same values because the new BUA vector is generated by adding one to the current BUA vector, say *X*, in a way similar to consecutively varying bits in a binary code.

The time complexity is a major index to compare the performance among the network-reliability related algorithms and the most efficient one always has the best time complexity [17, 18] theoretically.

**Property 2.** The number of all vectors generated from the BAT is $2^m$ with time complexity $O(2^{m+1})$.

**Proof.** Each vector has *m* coordinates and each coordinate is either 0 or 1. Hence, the number of all vectors generated from the BAT is $2^m$. The current vector needs to find out which coordinate *i* with value 0 such that any coordinate *j* with value 1 for all $i < j$ if it is a forward BAT, e.g., the proposed all-pairs BAT. or $i > j$ if it is a backward BAT, e.g., the traditional BAT. For example, as shown in Fig. 5, $X_1 = (0, 0, 0, 0, 0)$ takes only one comparison, i.e., the 1st coordinate, to find the first zero (in forward way) to have $X_2 = (1, 0, 0, 0, 0)$.

In the same way, the first zero coordinate is in the 2nd coordinate, i.e, $X_2$ needs two comparisons. The number of comparisons between two consecutive BUA vectors, e.g., $X_i$ and $X_i + 1$, can be found from the number of distinctive branch nodes in the tree shown in Fig. 5 as discussion in [25].

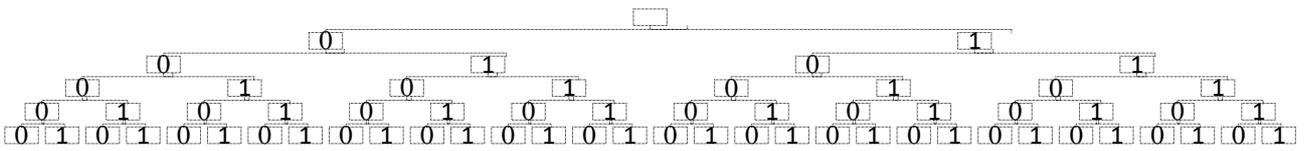

**Figure 5.** The number of comparisons between two consecutive BUA vectors.

Hence, the number of comparisons, which is also the total number of branch nodes, is

$$2 + 2^2 + 2^3 + \ldots + 2^m = 2(2^m - 1). \tag{15}$$

Thus, the complexity is $O(2^{m+1})$.

20The time complexity is $O(1\times 2^{m+1} + n\times 2^{m+1} + n^2\times 2^{m+1}) = O(n^2 2^{m+1})$ and $O(m\times 2^{m+1} + n\times 2^{m+1} + n^2\times 2^{m+1}) = O(n^2 2^{m+1})$ for the proposed all-pairs BAT no matter if $G(V, E, \mathbf{D})$ is homogeneous, where

$O(2^{m+1})$: the time complexity in finding all BUA vectors from STEPs 1–3

$O(n\times 2^{m+1})$: the time complexity to verify the connectivity of each BUA vector for any pair of nodes in STEP 4

$O(n^2\times 2^{m+1})$: the time complexity to calculate $R(s, t)$ for any pair of nodes $(s, t)$ in STEP 4

$O(1\times 2^{m+1})$: the time complexity to calculate $R_{s,t}(G)$ if $G(V, E, \mathbf{D})$ is homogeneous in STEP 5

$O(m\times 2^{m+1})$: the time complexity to calculate $R_{s,t}(G)$ if $G(V, E, \mathbf{D})$ is not homogeneous in STEP 6.

Hence, we have the following property about the time complexity of the proposed all-pairs BAT.

**Property 3.** The proposed all-pairs BAT finds all-pairs reliability with time complexity $O(n^2 2^{m+1})$.

There is no difference in the way to find each BUA in both traditional BAT and the proposed all-pairs BAT if the direction of each arc is ignored. However, the proposed CG-LSA outperforms the Floyd-Warshall algorithm [35, 36] and the all-pairs PLSA as discussion in Section 4.2. Without using the proposed CG-LSA, the time complexity is increased up to $O(n^3 2^{m+1})$. Hence, the proposed algorithm is efficient from the theoretical view.

### 5.3 Example

The bridge network depicted in Fig 1. is used to demonstrate the proposed all-pairs BAT because the bridge network can be solved using the all-pairs BAT by hand to avoid a complex and tedious calculations which is common NP-Hard obstacle of all network reliability problems.

There are five arcs in the bridge network and each related BUA vector needs to be 5-tuple. Each coordinate is either 0 or 1, i.e., there are $2^5 = 32$ different BUA vectors. Also, assume that each undirected arc with the success probability 0.9 and the failure probability 0.1, i.e., $\Pr(X) = P(|X|)$ for all obtained BUA vector $X$.

for arXiv

**Table 6.** Each BUA vector and its probability if it is connected.

| $i$ | $X_i$ | $\Pr(X_i)$ | (1, 2) | (1, 3) | (1, 4) | (2, 3) | (2, 4) | (3, 4) |
|---|---|---|---|---|---|---|---|---|
| 1  | (0, 0, 0, 0, 0) | 0.00000 | | | | | | |
| 2  | (1, 0, 0, 0, 0) | 0.00009 | 0.00009 | | | | | |
| 3  | (0, 1, 0, 0, 0) | 0.00009 | | 0.00009 | | | | |
| 4  | (1, 1, 0, 0, 0) | 0.00081 | 0.00081 | 0.00081 | | 0.00081 | | |
| 5  | (0, 0, 1, 0, 0) | 0.00009 | | | | 0.00009 | | |
| 6  | (1, 0, 1, 0, 0) | 0.00081 | 0.00081 | 0.00081 | | 0.00081 | | |
| 7  | (0, 1, 1, 0, 0) | 0.00081 | 0.00081 | 0.00081 | | 0.00081 | | |
| 8  | (1, 1, 1, 0, 0) | 0.00729 | 0.00729 | 0.00729 | | 0.00729 | | |
| 9  | (0, 0, 0, 1, 0) | 0.00009 | | | | | 0.00009 | |
| 10 | (1, 0, 0, 1, 0) | 0.00081 | 0.00081 | | 0.00081 | | 0.00081 | |
| 11 | (0, 1, 0, 1, 0) | 0.00081 | | 0.00081 | | | 0.00081 | |
| 12 | (1, 1, 0, 1, 0) | 0.00729 | 0.00729 | 0.00729 | 0.00729 | 0.00729 | 0.00729 | 0.00729 |
| 13 | (0, 0, 1, 1, 0) | 0.00081 | | | | 0.00081 | 0.00081 | 0.00081 |
| 14 | (1, 0, 1, 1, 0) | 0.00729 | 0.00729 | 0.00729 | 0.00729 | 0.00729 | 0.00729 | 0.00729 |
| 15 | (0, 1, 1, 1, 0) | 0.00729 | 0.00729 | 0.00729 | 0.00729 | 0.00729 | 0.00729 | 0.00729 |
| 16 | (1, 1, 1, 1, 0) | 0.06561 | 0.06561 | 0.06561 | 0.06561 | 0.06561 | 0.06561 | 0.06561 |
| 17 | (0, 0, 0, 0, 1) | 0.00009 | | | | | | 0.00009 |
| 18 | (1, 0, 0, 0, 1) | 0.00081 | 0.00081 | | | | | 0.00081 |
| 19 | (0, 1, 0, 0, 1) | 0.00081 | | 0.00081 | 0.00081 | | | 0.00081 |
| 20 | (1, 1, 0, 0, 1) | 0.00729 | 0.00729 | 0.00729 | 0.00729 | 0.00729 | 0.00729 | 0.00729 |
| 21 | (0, 0, 1, 0, 1) | 0.00081 | | | | 0.00081 | 0.00081 | 0.00081 |
| 22 | (1, 0, 1, 0, 1) | 0.00729 | 0.00729 | 0.00729 | 0.00729 | 0.00729 | 0.00729 | 0.00729 |
| 23 | (0, 1, 1, 0, 1) | 0.00729 | 0.00729 | 0.00729 | 0.00729 | 0.00729 | 0.00729 | 0.00729 |
| 24 | (1, 1, 1, 0, 1) | 0.06561 | 0.06561 | 0.06561 | 0.06561 | 0.06561 | 0.06561 | 0.06561 |
| 25 | (0, 0, 0, 1, 1) | 0.00081 | | | | 0.00081 | 0.00081 | 0.00081 |
| 26 | (1, 0, 0, 1, 1) | 0.00729 | 0.00729 | 0.00729 | 0.00729 | 0.00729 | 0.00729 | 0.00729 |
| 27 | (0, 1, 0, 1, 1) | 0.00729 | 0.00729 | 0.00729 | 0.00729 | 0.00729 | 0.00729 | 0.00729 |
| 28 | (1, 1, 0, 1, 1) | 0.06561 | 0.06561 | 0.06561 | 0.06561 | 0.06561 | 0.06561 | 0.06561 |
| 29 | (0, 0, 1, 1, 1) | 0.00729 | | | | 0.00729 | 0.00729 | 0.00729 |
| 30 | (1, 0, 1, 1, 1) | 0.06561 | 0.06561 | 0.06561 | 0.06561 | 0.06561 | 0.06561 | 0.06561 |
| 31 | (0, 1, 1, 1, 1) | 0.06561 | 0.06561 | 0.06561 | 0.06561 | 0.06561 | 0.06561 | 0.06561 |
| 32 | (1, 1, 1, 1, 1) | 0.59049 | 0.59049 | 0.59049 | 0.59049 | 0.59049 | 0.59049 | 0.59049 |
| | SUM | 1.00000 | 0.98829 | 0.98829 | 0.97848 | 0.99639 | 0.98829 | 0.98829 |

As most of BAT algorithms and the pseudocode presented in Section 5.2, the BUA vector $X$ is initialized to be zero vector, i.e., (0, 0, 0, 0, 0) which is obviously not connected for any pair of nodes.

After the initial BUA vector, all other BUA vector is obtained based on the forward binary addition also as shown in the pseudocode listed in Section 5.2, e.g., (1, 0, 0, 0, 0), (0, 1, 0, 0, 0), etc. Analogy, we have all the BUA vectors listed in Table 6. For easy recognition, $X_i$ is the BUA vector $X$ found in the $i$th iteration for $i = 1, 2, \ldots, 32$. Note that there is no need to record all BUA vectors because each BUA is discarded after it has tested the connectivity between any two nodes using the proposed CG-LSA.

for arXiv

In Table 6, these empty grids mean the related pair of nodes are connected; otherwise, the grid lists the related probability, e.g., 0.00009 in the column under title (1, 2) and the row under $i = 2$ means that the corresponding $X_2 = (1, 0, 0, 0, 0)$ is (1, 2)-connected with probability 0.00009. Details regarding how to verify the state vectors are presented in Section 4.2. The last row is the summations of all probabilities that the related pair is connected, e.g, $R_{1,2}(G) = 0.98829$.

Hence, the lower triangular of the symmetry reliability matrix is:

$$\mathbf{R} = \begin{bmatrix} 0 & & & \\ 0.98829 & 0 & & \\ 0.98829 & 0.99639 & 0 & \\ 0.97848 & 0.98829 & 0.98829 & 0 \end{bmatrix}, \quad (16)$$

where the element in the $i^{th}$ row and the $j$th column is $R_{i,j}(G)$, e.g., $R_{4,2}(G) = 0.98829$.

**5.4 Computation Experiments**

The performance and effectiveness of the proposed all-pairs BAT is validated by testing on 20 undirected binary-state benchmark networks shown in Fig. 6 [20, 21, 25]. These popular binary-state benchmark networks are changed to undirected networks to meet the requirement of the proposed all-pairs BAT, e.g., Fig. 6(6). Also, each arc is set to be homogeneous with the reliability 0.9.

The testing environments all adapted from [21] which compared the traditional BAT with QIE [31], including the code are programmed in DEV C++ 5.11, the platform is 64-bit Windows 10, the computer hardware is Intel Core i7-6650U CPU @ 2.20GHz 2.21GHz notebook and 16 GB RAM.

Table 7 provides the experimental results of the proposed all-pairs BAT. In Table 7, the notations $N_{OLD}$, $N_{BAT}$, $T_{OLD}$, $T_{BAT}$, $N^{\#}$, $\mathbf{C}$, and $\mathbf{R}$ are the total number of vectors obtained from the traditional BAT [25] and the proposed all-pairs BAT, the average number of connected vectors, the total running time for the traditional BAT [25] and the proposed all-pairs BAT, the connection matrix, and the reliability matrix, respectively. Note that these empty grids in columns $N_{OLD}$ and $T_{OLD}$ mean that the traditional BAT is failed to calculate the one-pair reliability within 10 hours.

From the running time $T_{BAT}$ listed in Table 7, the proposed all-pairs BAT can solve all

for arXiv

benchmark problems within 30 seconds. Moreover, the running time of these networks, e.g., Fig. 6(1)-6(5), 6(8)-6(12), with arc number $m < 14$ are all zeros. Apparently, the NP-Hard characteristic, i.e., the running time increased exponentially with the number of arcs, start to stronger from $m = 14$. For example, the running times are 153.06, 619.22 (almost $4 = 2^2$ times of 153.06), and 1376.18 (around double of 619.22) corresponding to $m = 27, 29 (= 27 + 2), 30 = (29 + 1)$ for Fig. 6(17), 6(14), and 6(13).

The traditional BAT is unable to calculate a one-pair reliability of network in Fig. 6(13), 6(14), 6(17), 6(18), 6(19), and 6(20) within 10 hours [25]. Conversely, the proposed all-pairs BAT can solve all-pairs reliability, i.e., the $n(n-1)/2$ pairs, for each network within 30 minutes. The main reason for that is the proposed all-pairs BAT can solve the problems directly without needing to change each undirected arc into two directed arcs which square of the running time than that of the proposed all-pairs BAT. Another reason is the implementation of the CG-LSA can reduce the running time down to $1/n$ of the original one to further reduce the computational burden.

Another interesting phenomenon is the number of disconnected vectors is less than connected vectors for almost all cases except Fig. 6(5), 6(14), 6(17), 6(19), and 6(20). For example, in Fig. 6(1), the total number of vectors generated in the BAT is 32 and average number of connected vectors is

$$[21 + (21+23) + (16+21+21)]/6 = 20.5, \tag{17}$$

i.e., the average number of disconnected vectors is 32 – 20.5 = 11.5. Also, among 6 distinctive pairs, e.g., (1, 2), (1, 3), (1, 4), (2, 3), (2, 4), and (3, 4), 5 pairs have more than 16 connected vectors and one pair has exactly 16 connected vectors. From the average number and the number of pairs of the connected vectors, it is more efficient to calculate the binary-state network reliability based on the disconnected vectors, e.g., the minimal cuts, than the connected vectors, e.g., the minimal paths. This important observation further confirms these conclusions obtained in [23, 25] that cut-based algorithms are more efficient that path-based algorithms and can be used to calculate the network reliability by developing disconnected-vector algorithms.

for arXiv

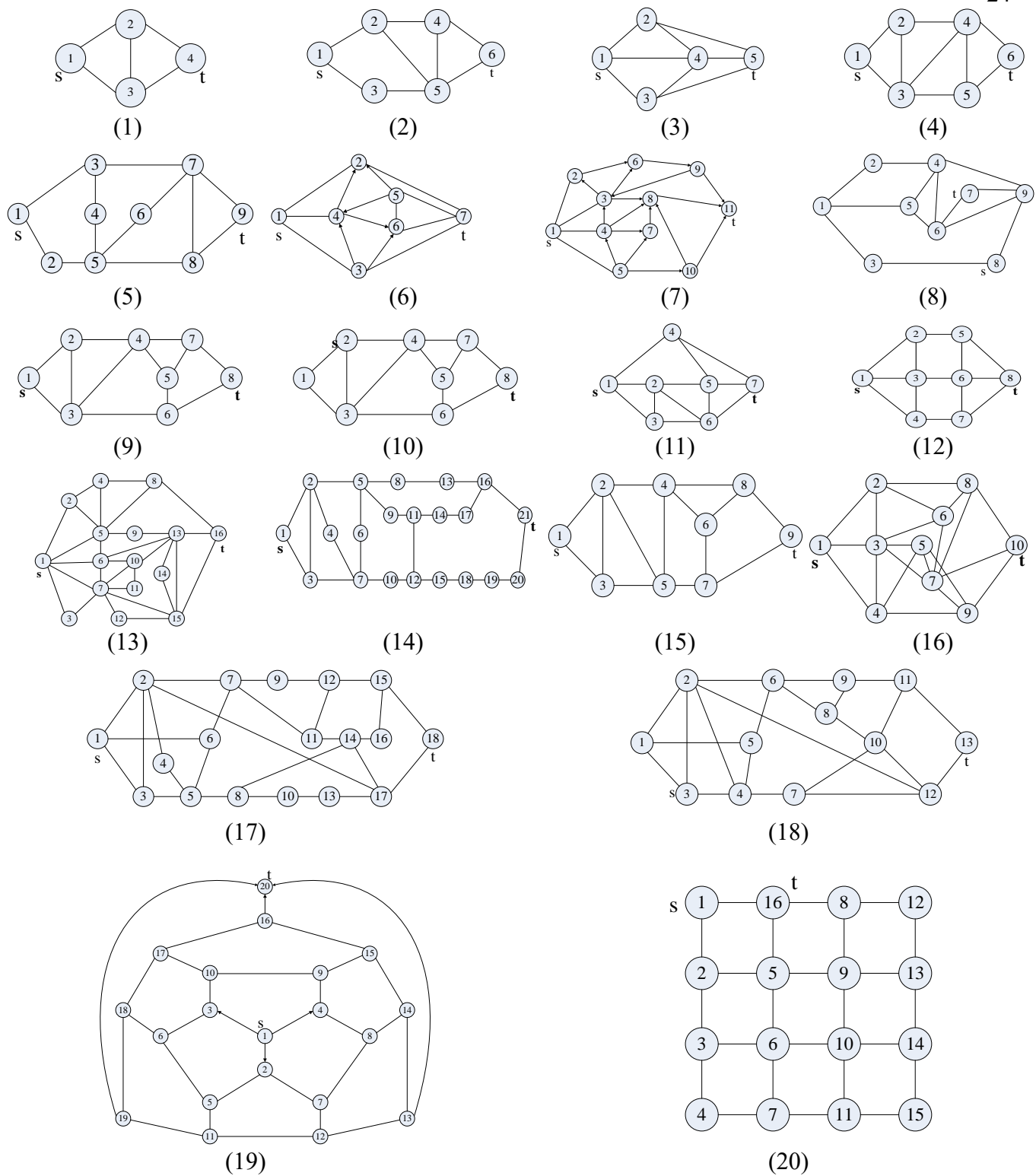

**Figure 6.** 20 binary-state benchmark networks.

From the above, the proposed all-pairs BAT is superior than the traditional BAT [25] which outperforms the reliability algorithms based on the minimum cuts [23], minimum paths [24], SDP [32], or IET [31].

for arXiv

**Table 7.** The experimental results for the proposed all-pairs BAT.

| Fig. 6 | $n$ | $m$ | $N_{old}$ | $T_{old}$ | $N_{BAT}$ | $T_{BAT}$ | $N^{\#}$ | **C** and **R** |
|---|---|---|---|---|---|---|---|---|
| (1) | 4 | 5 | 32 | 0.00000000 | 32 | 0.00000000000 | 20.5 | Appendix Eq. (1) |
| (2) | 6 | 8 | 1456 | 0.00000000 | 256 | 0.00000000000 | 142.1 | Appendix Eq. (2) |
| (3) | 5 | 8 | 688 | 0.00000000 | 256 | 0.00000000000 | 188.0 | Appendix Eq. (3) |
| (4) | 6 | 9 | 6560 | 0.00200000 | 512 | 0.00000000000 | 322.0 | Appendix Eq. (4) |
| (5) | 9 | 12 | 277760 | 0.08400000 | 4096 | 0.00000000000 | 1908.6 | Appendix Eq. (5) |
| (6) | 7 | 14 | 20096 | 0.00000000 | 16384 | 0.01499999970 | 13272.5 | Appendix Eq. (6) |
| (7) | 11 | 21 | 991522 | 0.23199999 | 2097152 | 1.41400003430 | 1474792.8 | Appendix Eq. (7) |
| (8) | 9 | 13 | 1574912 | 0.28600000 | 8192 | 0.00000000000 | 4153.8 | Appendix Eq. (8) |
| (9) | 8 | 12 | 335104 | 0.08200000 | 4096 | 0.00000000000 | 2318.9 | Appendix Eq. (9) |
| (10) | 8 | 12 | 4877312 | 1.35099995 | 4096 | 0.00000000000 | 2318.9 | Appendix Eq. (10) |
| (11) | 7 | 12 | 158208 | 0.02400000 | 4096 | 0.00000000000 | 2889.0 | Appendix Eq. (11) |
| (12) | 8 | 13 | 524288 | 0.10500000 | 8192 | 0.00000000000 | 5174.5 | Appendix Eq. (12) |
| (13) | 16 | 30 | | | 1073741824 | 1376.18005371090 | 687075488.2 | Appendix Eq. (13) |
| (14) | 21 | 29 | | | 536870912 | 619.22998046880 | 13671437.8 | Appendix Eq. (14) |
| (15) | 9 | 14 | 4708352 | 1.11399996 | 16384 | 0.00000000000 | 9100.2 | Appendix Eq. (15) |
| (16) | 10 | 21 | 44719538176 | 6950.24023438 | 2097152 | 1.46099996570 | 1661558.5 | Appendix Eq. (16) |
| (17) | 18 | 27 | | | 134217728 | 153.60000610350 | 56023006.1 | Appendix Eq. (17) |
| (18) | 13 | 22 | | | 4194304 | 3.32999992370 | 2401803.9 | Appendix Eq. (18) |
| (19) | 20 | 30 | | | 1073741824 | 1455.59997558590 | 429533547.2 | Appendix Eq. (19) |
| (20) | 16 | 24 | | | 16777216 | 15.07199954990 | 7279993.1 | Appendix Eq. (20) |

## 6. CONCLUSIONS

Reliability is the success probability that the network is still functioning and it plays significant roles in assessing the performances of binary-state networks which are the basis of many network applications. This paper presents a novel all-pairs BAT based on the undirected BAT, CG-LSA, and the reliability pre-calculation method to solve the homogeneity-Arc binary-state undirected network reliability problems.

From the complete experimental test, the proposed all-pairs BAT is able to calculate each pair of reliability for all 20 well-known binary-state undirected networks within 30min and is a big improvement compared to other existing-known algorithms which are all impossible to calculate the one-pair reliability for all 20 benchmark problems within 10hrs. These outcomes also confirm that the time complexity of the proposed all-pairs BAT is better than existing-known algorithms to find all combinations of any two nodes as discussed in Section 5.1.

In future works, the proposed all-pairs BAT will be further improved and extended to all-pairs reliability of the multi-state flow network, multi-state information network, and multi-commodity

for arXiv

network, etc., with directed arcs, heterogeneous arcs, and unreliable nodes.

**ACKNOWLEDGEMENTS**

This research was supported in part by the Ministry of Science and Technology, R.O.C. under grant MOST 107-2221-E-007-072-MY3 and MOST 104-2221-E-007-061-MY3. This article was once submitted to arXiv as a temporary submission that was just for reference and did not provide the copyright.

for arXiv

for arXiv

29[30] W. C. Yeh, "One-Batch Preempt Multi-State Multi-Rework Network Reliability Problem," *Reliability Engineering & System Safety*, 2020, under 1st revision, 2021/5.

[31] Z. Hao, W. C. Yeh, J. Wang, G. G. Wang, and B. Sun, "A Quick Inclusion-Exclusion Technique," *Information Sciences*, vol. 486, pp. 20-30, 2019.

[32] M. J. Zuo, Z. Tian, and H. Z. Huang, "An efficient method for reliability evaluation of multistate networks given all minimal path vectors", *IIE Transactions*, vol. 39, no. 8, pp. 811-817, 2007.

[33] G. Bai, M. J. Zuo, and Z. Tian, "Search for all *d*-MPs for all *d* levels in multistate two-terminal networks." *Reliability Engineering & System Safety*, vol. 142, pp. 300-309, 2015.

[34] W. C. Yeh, "A revised layered-network algorithm to search for all d-minpaths of a limited-flow acyclic network," *IEEE Transactions on Reliability*, vol. 47, no. 4, pp. 436-442, 1998.

[35] C. Papadimitriou, and M. Sideri, "On the Floyd–Warshall algorithm for logic programs," *The Journal of Logic Programming*, vol. 41, no. 1, pp. 129–137, October 1999.

[36] A. Aini, and A. Salehipour, "Speeding up the Floyd–Warshall algorithm for the cycled shortest path problem," *Applied Mathematics Letters*, vol. 25, no. 1, pp. 1–5, January 2012.
for arXiv

for arXiv

**APENDIX**

$$C(G)=\begin{bmatrix} 0 & & & \\ 21 & 0 & & \\ 21 & 23 & 0 & \\ 16 & 21 & 21 & 0 \end{bmatrix}, R(G)=\begin{bmatrix} 0 & & & \\ 0.98829 & 0 & & \\ 0.98829 & 0.99639 & 0 & \\ 0.97848 & 0.98829 & 0.98829 & 0 \end{bmatrix} \quad (1)$$

$$C(G)=\begin{bmatrix} 0 & & & & \\ 149 & 0 & & & \\ 149 & 127 & 0 & & \\ 112 & 173 & 116 & 0 & \\ 127 & 179 & 149 & 187 & 0 \\ 91 & 134 & 101 & 169 & 169 & 0 \end{bmatrix}, R(G)=\begin{bmatrix} 0 & & & & & \\ 0.98005149 & 0 & & & & \\ 0.98005149 & 0.97899759 & 0 & & & \\ 0.97732008 & 0.99544149 & 0.97804908 & 0 & & \\ 0.97899759 & 0.99682659 & 0.98005149 & 0.99763659 & 0 & \\ 0.96842547 & 0.98622198 & 0.96930837 & 0.98888049 & 0.98888049 & 0 \end{bmatrix} \quad (2)$$

$$C(G)=\begin{bmatrix} 0 & & & & \\ 187 & 0 & & & \\ 187 & 172 & 0 & & \\ 197 & 197 & 197 & 0 & \\ 172 & 187 & 187 & 197 & 0 \end{bmatrix}, R(G)=\begin{bmatrix} 0 & & & & \\ 0.99781803 & 0 & & & \\ 0.99781803 & 0.99763164 & 0 & & \\ 0.99870093 & 0.99870093 & 0.99870093 & 0 & \\ 0.99763164 & 0.99781803 & 0.99781803 & 0.99870093 & 0 \end{bmatrix} \quad (3)$$

$$C(G)=\begin{bmatrix} 0 & & & & & \\ 341 & 0 & & & & \\ 341 & 383 & 0 & & & \\ 286 & 361 & 391 & 0 & & \\ 253 & 304 & 361 & 383 & 0 & \\ 205 & 253 & 286 & 341 & 341 & 0 \end{bmatrix}, R(G)=\begin{bmatrix} 0 & & & & & \\ 0.989005149 & 0 & & & & \\ 0.989005149 & 0.997899759 & 0 & & & \\ 0.987856398 & 0.996951249 & 0.998628759 & 0 & & \\ 0.986165037 & 0.995212008 & 0.996951249 & 0.997899759 & 0 & \\ 0.977184405 & 0.986165037 & 0.987856398 & 0.989005149 & 0.989005149 & 0 \end{bmatrix} \quad (4)$$

$$C(G)=\begin{bmatrix} 0 & & & & & & & & \\ 2269 & 0 & & & & & & & \\ 2269 & 1687 & 0 & & & & & & \\ 1532 & 1532 & 2393 & 0 & & & & & \\ 1687 & 2269 & 2059 & 2393 & 0 & & & & \\ 1235 & 1457 & 1712 & 1627 & 2489 & 0 & & & \\ 1522 & 1510 & 2389 & 1832 & 2347 & 2489 & 0 & & \\ 1321 & 1507 & 1882 & 1709 & 2533 & 2072 & 2851 & 0 & \\ 1085 & 1151 & 1631 & 1351 & 1862 & 1741 & 2657 & 2657 & 0 \end{bmatrix}, R(G)=\begin{bmatrix} 0 & & & & & & & & \\ 0.979152829 & 0 & & & & & & & \\ 0.979152829 & 0.977100417 & 0 & & & & & & \\ 0.968467853 & 0.968467853 & 0.987072612 & 0 & & & & & \\ 0.977100417 & 0.979152829 & 0.993819959 & 0.987072612 & 0 & & & & \\ 0.966519709 & 0.968194587 & 0.983414468 & 0.97620505 & 0.988457712 & 0 & & & \\ 0.975332489 & 0.976662031 & 0.99274066 & 0.984924227 & 0.996744059 & 0.988457712 & 0 & & \\ 0.973772173 & 0.975210199 & 0.991042293 & 0.98340437 & 0.99537235 & 0.986601737 & 0.997624538 & 0 & \\ 0.964855123 & 0.966225126 & 0.982021902 & 0.974371467 & 0.986147131 & 0.97770359 & 0.988874781 & 0.988874781 & 0 \end{bmatrix} \quad (5)$$

$$C(G)=\begin{bmatrix} 0 & & & & & & \\ 12836 & 0 & & & & & \\ 13012 & 13472 & 0 & & & & \\ 13012 & 13472 & 13912 & 0 & & & \\ 12022 & 13448 & 14020 & 12976 & 0 & & \\ 12068 & 12988 & 14132 & 13436 & 13778 & 0 & \\ 12022 & 13448 & 14020 & 12976 & 13894 & 13778 & 0 \end{bmatrix}, R(G)=\begin{bmatrix} 0 & & & & & & \\ 0.998886551 & 0 & & & & & \\ 0.998961472 & 0.999864854 & 0 & & & & \\ 0.998889107 & 0.999777212 & 0.99986795 & 0 & & & \\ 0.998865481 & 0.999776485 & 0.99988297 & 0.999773295 & 0 & & \\ 0.998866341 & 0.999773367 & 0.99988401 & 0.999776408 & 0.999794921 & 0 & \\ 0.998865481 & 0.999776485 & 0.99988297 & 0.999773295 & 0.999795967 & 0.999794921 & 0 \end{bmatrix} \quad (6)$$

for arXiv

$$C(G) = \begin{bmatrix} 0 & & & & & & & & & \\ 1578572 & 0 & & & & & & & & \\ 1754846 & 1657736 & 0 & & & & & & & \\ 1743548 & 1482374 & 1770320 & 0 & & & & & & \\ 1652942 & 1370000 & 1610876 & 1752344 & 0 & & & & & \\ 1427386 & 1526752 & 1625296 & 1413772 & 1297744 & 0 & & & & \\ 1492382 & 1273262 & 1520870 & 1680026 & 1628570 & 1219804 & 0 & & & \\ 1611598 & 1421086 & 1722310 & 1775134 & 1646494 & 1383524 & 1637686 & 0 & & \\ 1379084 & 1338176 & 1584536 & 1406330 & 1300460 & 1497472 & 1228238 & 1420198 & 0 & \\ 1386986 & 1206290 & 1444520 & 1497548 & 1532612 & 1179010 & 1377392 & 1615648 & 1246418 & 0 \\ 1332596 & 1203404 & 1443746 & 1428110 & 1364954 & 1226086 & 1296110 & 1572544 & 1424852 & 1503032 & 0 \end{bmatrix},$$

$$R(G) = \begin{bmatrix} 0 & & & & & & & & & \\ 0.998788186 & 0 & & & & & & & & \\ 0.999873607 & 0.998888876 & 0 & & & & & & & \\ 0.999864856 & 0.998854004 & 0.999960477 & 0 & & & & & & \\ 0.999756165 & 0.99874336 & 0.999849487 & 0.999866582 & 0 & & & & & \\ 0.9986781 & 0.997901759 & 0.998800046 & 0.998763079 & 0.998652283 & 0 & & & & \\ 0.998861937 & 0.997852274 & 0.99895765 & 0.998976633 & 0.998886728 & 0.997761492 & 0 & & & \\ 0.999849665 & 0.998841374 & 0.999948075 & 0.999960709 & 0.999851824 & 0.998750914 & 0.998964448 & 0 & & \\ 0.998667861 & 0.997701863 & 0.998789788 & 0.998756501 & 0.998646052 & 0.997819944 & 0.997755311 & 0.998746753 & 0 & \\ 0.998752387 & 0.997745103 & 0.9988504 & 0.998862889 & 0.998777175 & 0.997656384 & 0.997867615 & 0.998878073 & 0.997669924 & 0 \\ 0.998663255 & 0.997659968 & 0.998763717 & 0.998771894 & 0.998665303 & 0.997589128 & 0.997774794 & 0.998787502 & 0.997789947 & 0.997898376 & 0 \end{bmatrix} \quad (7)$$

$$C(G) = \begin{bmatrix} 0 & & & & & & & \\ 3227 & 0 & & & & & & \\ 2600 & 4889 & 0 & & & & & \\ 4489 & 4489 & 2932 & 0 & & & & \\ 3512 & 4427 & 4889 & 3100 & 0 & & & \\ 3161 & 4991 & 4187 & 3187 & 5861 & 0 & & \\ 3595 & 4042 & 3964 & 2981 & 6217 & 5707 & 0 & \\ 4489 & 3775 & 3727 & 3227 & 5869 & 4900 & 6191 & 0 \\ 3076 & 2971 & 2923 & 2360 & 4594 & 4033 & 5477 & 5477 & 0 \end{bmatrix}, \quad R(G) = \begin{bmatrix} 0 & & & & & & & \\ 0.976961994 & 0 & & & & & & \\ 0.968116594 & 0.987233602 & 0 & & & & & \\ 0.97908726 & 0.97908726 & 0.968377131 & 0 & & & & \\ 0.978595403 & 0.994159827 & 0.987233602 & 0.976996107 & 0 & & & \\ 0.977651197 & 0.993828844 & 0.986230827 & 0.976364702 & 0.998307705 & 0 & & \\ 0.978683951 & 0.993879646 & 0.986750039 & 0.976900926 & 0.999304045 & 0.998076363 & 0 & \\ 0.97908726 & 0.993600587 & 0.986449097 & 0.976961994 & 0.998975483 & 0.997702625 & 0.999248532 & 0 \\ 0.969111795 & 0.98381834 & 0.976749135 & 0.96717706 & 0.989164232 & 0.987926414 & 0.989644041 & 0.989644041 & 0 \end{bmatrix} \quad (8)$$

$$C(G) = \begin{bmatrix} 0 & & & & & & & \\ 2714 & 0 & & & & & & \\ 2714 & 3022 & 0 & & & & & \\ 2204 & 2818 & 2974 & 0 & & & & \\ 1714 & 2123 & 2381 & 2876 & 0 & & & \\ 1786 & 2096 & 2594 & 2462 & 2704 & 0 & & \\ 1618 & 2021 & 2231 & 2792 & 2836 & 2362 & 0 & \\ 1309 & 1583 & 1856 & 2021 & 2131 & 2494 & 2494 & 0 \end{bmatrix}, \quad R(G) = \begin{bmatrix} 0 & & & & & & & \\ 0.988996344 & 0 & & & & & & \\ 0.988996344 & 0.997881171 & 0 & & & & & \\ 0.987740956 & 0.996844612 & 0.998501469 & 0 & & & & \\ 0.985585123 & 0.994649427 & 0.996341647 & 0.997583447 & 0 & & & \\ 0.984974424 & 0.993999763 & 0.995757592 & 0.996564669 & 0.996688081 & 0 & & \\ 0.984776866 & 0.9938352 & 0.99552311 & 0.996782932 & 0.996868298 & 0.995698544 & 0 & \\ 0.975115897 & 0.984068153 & 0.985773961 & 0.986797498 & 0.986900885 & 0.987962468 & 0.987962468 & 0 \end{bmatrix} \quad (9)$$

for arXiv

$$C(G)=\begin{bmatrix} 0 \\ 2714 & 0 \\ 3022 & 2714 & 0 \\ 2818 & 2204 & 2974 & 0 \\ 2123 & 1714 & 2381 & 2876 & 0 \\ 2096 & 1786 & 2594 & 2462 & 2704 & 0 \\ 2021 & 1618 & 2231 & 2792 & 2836 & 2362 & 0 \\ 1583 & 1309 & 1856 & 2021 & 2131 & 2494 & 2494 & 0 \end{bmatrix},\ R(G)=\begin{bmatrix} 0 \\ 0.988996344 & 0 \\ 0.997881171 & 0.988996344 & 0 \\ 0.996844612 & 0.987740956 & 0.998501469 & 0 \\ 0.994649427 & 0.985585123 & 0.996341647 & 0.997583447 & 0 \\ 0.993999763 & 0.984974424 & 0.995757592 & 0.996564669 & 0.996688081 & 0 \\ 0.9938352 & 0.984776866 & 0.99552311 & 0.996782932 & 0.996868298 & 0.995698544 & 0 \\ 0.984068153 & 0.975115897 & 0.985773961 & 0.986797498 & 0.986900885 & 0.987962468 & 0.987962468 & 0 \end{bmatrix} \quad (10)$$

$$C(G)=\begin{bmatrix} 0 \\ 2776 & 0 \\ 3048 & 2712 & 0 \\ 2952 & 2472 & 3152 & 0 \\ 2712 & 3048 & 3146 & 2746 & 0 \\ 2758 & 2758 & 3258 & 3036 & 3258 & 0 \\ 2472 & 2952 & 2746 & 2480 & 3152 & 3036 & 0 \end{bmatrix},\ R(G)=\begin{bmatrix} 0 \\ 0.997621376 & 0 \\ 0.998685243 & 0.998472563 & 0 \\ 0.997876986 & 0.997493674 & 0.998782135 & 0 \\ 0.998472563 & 0.998685243 & 0.999550777 & 0.998551249 & 0 \\ 0.998558519 & 0.998558519 & 0.999654755 & 0.998677501 & 0.999654755 & 0 \\ 0.997493674 & 0.997876986 & 0.998551249 & 0.997557919 & 0.998782135 & 0.998677501 & 0 \end{bmatrix} \quad (11)$$

$$C(G)=\begin{bmatrix} 0 \\ 5745 & 0 \\ 6159 & 5967 & 0 \\ 5745 & 4932 & 5967 & 0 \\ 4404 & 5265 & 4959 & 4349 & 0 \\ 4722 & 4959 & 5635 & 4959 & 5967 & 0 \\ 4404 & 4349 & 4959 & 5265 & 4932 & 5967 & 0 \\ 4096 & 4404 & 4722 & 4404 & 5745 & 6159 & 5745 & 0 \end{bmatrix},\ R(G)=\begin{bmatrix} 0 \\ 0.997703592 & 0 \\ 0.998679809 & 0.998583121 & 0 \\ 0.997703592 & 0.997400814 & 0.998583121 & 0 \\ 0.996338212 & 0.996649692 & 0.997403419 & 0.996406468 & 0 \\ 0.997298567 & 0.997403419 & 0.998386153 & 0.997403419 & 0.998583121 & 0 \\ 0.996338212 & 0.996406468 & 0.997403419 & 0.996649692 & 0.997400814 & 0.998583121 & 0 \\ 0.996217493 & 0.996338212 & 0.997298567 & 0.996338212 & 0.997703592 & 0.998679809 & 0.997703592 & 0 \end{bmatrix} \quad (12)$$

$$C(G)=\begin{bmatrix}
0 \\
809248142 & 0 \\
750661634 & 583623850 & 0 \\
733859362 & 782146274 & 538361666 & 0 \\
898308412 & 847925894 & 660179420 & 830553742 & 0 \\
915253334 & 755347534 & 703374634 & 714403670 & 885344534 & 0 \\
909807622 & 734165402 & 750661634 & 690151720 & 847782442 & 944953922 & 0 \\
715198462 & 686085944 & 533630150 & 765817762 & 807104614 & 715557314 & 696466822 & 0 \\
658310785 & 588735275 & 499090763 & 572089741 & 722413174 & 680496743 & 662020894 & 573449527 & 0 \\
821119162 & 673819676 & 654573338 & 638260408 & 785479948 & 895097420 & 909754630 & 647621056 & 628451221 & 0 \\
654595418 & 532385980 & 531567634 & 502271114 & 617548802 & 695763460 & 750643970 & 508171388 & 487870619 & 750643970 & 0 \\
646218994 & 528943460 & 525278066 & 502599544 & 614192896 & 678129458 & 742377898 & 517490116 & 492936385 & 650935984 & 527019260 & 0 \\
834611728 & 701126456 & 653955716 & 674478886 & 825062242 & 905871560 & 894962680 & 702295372 & 722413174 & 868786240 & 666821162 & 683503522 & 0 \\
616607776 & 515622242 & 487585736 & 495941650 & 604686556 & 662270264 & 672906802 & 518982010 & 515338225 & 634401394 & 494247620 & 539109988 & 750855274 & 0 \\
796468960 & 662864150 & 635864324 & 637497856 & 774454786 & 845768888 & 884956414 & 670735042 & 640239181 & 809166394 & 640603394 & 742377898 & 910388542 & 750855274 & 0 \\
710478920 & 617227390 & 553123408 & 621600842 & 724817972 & 749441356 & 752625518 & 723523454 & 586614767 & 707402978 & 551984308 & 590335088 & 819384854 & 614707544 & 806760722 & 0
\end{bmatrix},$$

$$R(G)=\begin{bmatrix}
0 \\
0.998859842 & 0 \\
0.989987325 & 0.988877271 & 0 \\
0.99875777 & 0.997908974 & 0.988778024 & 0 \\
0.999952291 & 0.998895276 & 0.989960756 & 0.998800684 & 0 \\
0.999964566 & 0.998862336 & 0.989977148 & 0.998763909 & 0.999958745 & 0 \\
0.999973242 & 0.998870878 & 0.989987325 & 0.998772444 & 0.999967279 & 0.999988117 & 0 \\
0.998758397 & 0.997718201 & 0.988779178 & 0.997828787 & 0.998798161 & 0.998765542 & 0.998774149 & 0 \\
0.989960269 & 0.988891645 & 0.980070873 & 0.98879616 & 0.989981229 & 0.989970876 & 0.989979756 & 0.988796124 & 0 \\
0.999783641 & 0.998681571 & 0.989798861 & 0.998583166 & 0.999777774 & 0.999798989 & 0.999808866 & 0.998584883 & 0.989792297 & 0 \\
0.989883731 & 0.988792532 & 0.979998195 & 0.988695096 & 0.989877875 & 0.989898691 & 0.989909463 & 0.98869679 & 0.979991199 & 0.989909463 & 0 \\
0.989953084 & 0.988862018 & 0.980066703 & 0.988764656 & 0.989947453 & 0.989967844 & 0.989978511 & 0.988767 & 0.980062135 & 0.989790093 & 0.979989491 & 0 \\
0.999965741 & 0.998863929 & 0.989978721 & 0.998765618 & 0.999960373 & 0.999980714 & 0.999990119 & 0.998768076 & 0.989981229 & 0.999800919 & 0.989900637 & 0.989974372 & 0 \\
0.989949339 & 0.988858546 & 0.980062412 & 0.988761245 & 0.989944002 & 0.989964147 & 0.989973586 & 0.988763962 & 0.980062833 & 0.989786126 & 0.979985091 & 0.980077566 & 0.989979231 & 0 \\
0.999930026 & 0.998828218 & 0.989943477 & 0.998729961 & 0.999924613 & 0.999944968 & 0.999954634 & 0.998732996 & 0.989941824 & 0.999765126 & 0.989865355 & 0.989978511 & 0.999956154 & 0.989979231 & 0 \\
0.998865938 & 0.997770565 & 0.988889543 & 0.997690472 & 0.998864492 & 0.998880198 & 0.998889564 & 0.997881472 & 0.988890013 & 0.998700404 & 0.988811104 & 0.988888207 & 0.99889142 & 0.988889093 & 0.998862454 & 0
\end{bmatrix} \quad (13)$$

$$C(G)=\begin{bmatrix}
0 \\
43423964 & 0 \\
43423964 & 46385812 & 0 \\
28271788 & 40382588 & 33658784 & 0 \\
25675172 & 38022028 & 29173720 & 26316224 & 0 \\
21569092 & 29245184 & 27266060 & 25946644 & 38039420 & 0 \\
29857064 & 37261684 & 40959052 & 40382588 & 30232180 & 38039420 & 0 \\
13013828 & 19252600 & 14807236 & 13366796 & 33884836 & 19263536 & 15416812 & 0 \\
14034694 & 20493170 & 16246676 & 14802448 & 34721984 & 20534410 & 17907422 & 18063242 & 0 \\
15885092 & 20093944 & 21525220 & 21115244 & 17974900 & 20478464 & 34515340 & 9452758 & 14852552 & 0 \\
9747029 & 13670743 & 11855566 & 11076608 & 20279872 & 13760627 & 15052057 & 11638255 & 34721984 & 19660759 & 0 \\
10157852 & 13381888 & 13230100 & 12759884 & 15185524 & 13569800 & 19659940 & 8476747 & 19694909 & 34515340 & 34577974 & 0 \\
6931598 & 10213246 & 7929304 & 7178756 & 17768428 & 10223582 & 8406958 & 33884836 & 10591820 & 5742700 & 8978647 & 6113995 & 0 \\
5428781 & 7656550 & 6560077 & 6109367 & 11585257 & 7701992 & 8186959 & 8817400 & 18058424 & 10250671 & 33923380 & 17698399 & 10617058 & 0 \\
5164444 & 6816863 & 6713051 & 6468763 & 7811957 & 6910957 & 9934802 & 4627280 & 10013011 & 17334734 & 17491319 & 33647564 & 3819362 & 9223652 & 0 \\
4387367 & 6384289 & 5100958 & 4658084 & 10701436 & 6399581 & 5697691 & 17768428 & 8572007 & 4989697 & 10971244 & 6915892 & 33884836 & 17884060 & 4984244 & 0 \\
3897449 & 5574706 & 4630189 & 4275503 & 8848561 & 5598836 & 5514283 & 10568878 & 10587002 & 6058939 & 17884060 & 9794947 & 17733112 & 33923380 & 5650202 & 33923380 & 0 \\
2767532 & 3680029 & 3570073 & 3428489 & 4371751 & 3727511 & 5199646 & 3117178 & 5367167 & 8850082 & 9194569 & 17056612 & 3465616 & 5400910 & 33647564 & 5591284 & 4371400 & 0 \\
1768660 & 2402969 & 2229677 & 2118925 & 3144803 & 2427739 & 3086894 & 3191390 & 3434455 & 4819058 & 5539349 & 9040532 & 4875884 & 4318802 & 17056612 & 9040532 & 5319140 & 33647564 & 0 \\
1668392 & 2347153 & 2021665 & 1885289 & 3517639 & 2361755 & 2540170 & 4887022 & 3248519 & 3226150 & 4698049 & 5591284 & 8755300 & 5436274 & 9040532 & 17056612 & 8967292 & 17056612 & 33647564 & 0 \\
2416594 & 3484673 & 2842031 & 2610763 & 5676677 & 3496567 & 3286112 & 9051890 & 4716391 & 3274904 & 6250019 & 4984244 & 17043572 & 9312062 & 5591284 & 33647564 & 17139932 & 9040532 & 17056612 & 33647564 & 0
\end{bmatrix}$$

$$R(G)=\begin{bmatrix}
0 \\
0.9888716 & 0 \\
0.9888716 & 0.997617822 & 0 \\
0.977682326 & 0.988435927 & 0.986580305 & 0 \\
0.981796181 & 0.992516093 & 0.990810617 & 0.983167722 & 0 \\
0.97424479 & 0.984796067 & 0.983276378 & 0.976077852 & 0.986942032 & 0 \\
0.986105421 & 0.996698068 & 0.995333821 & 0.988435927 & 0.993544291 & 0.986942032 & 0 \\
0.950216518 & 0.96058307 & 0.958949678 & 0.951591115 & 0.967300592 & 0.95498143 & 0.961691228 & 0 \\
0.963618897 & 0.974122802 & 0.972484153 & 0.965061515 & 0.980391613 & 0.968229516 & 0.975362703 & 0.9518859 & 0 \\
0.964882703 & 0.975300837 & 0.973859324 & 0.966871921 & 0.975481605 & 0.96702207 & 0.977840315 & 0.947167446 & 0.963667076 & 0 \\
0.96406138 & 0.974552033 & 0.97294888 & 0.965603791 & 0.979715627 & 0.968222745 & 0.976029091 & 0.954893875 & 0.980391613 & 0.970506915 & 0 \\
0.962067463 & 0.972509958 & 0.970963134 & 0.96375162 & 0.976041635 & 0.965560733 & 0.974329554 & 0.950758776 & 0.970400799 & 0.977840315 & 0.983534636 & 0 \\
0.935814996 & 0.946015439 & 0.944424881 & 0.937218235 & 0.952079028 & 0.940282058 & 0.947225134 & 0.967300592 & 0.940624663 & 0.936015224 & 0.947394571 & 0.942723242 & 0 \\
0.938454747 & 0.948669361 & 0.947103593 & 0.939942 & 0.953854991 & 0.942569968 & 0.95007348 & 0.936836083 & 0.951814038 & 0.943819762 & 0.968187769 & 0.955620588 & 0.936868286 & 0 \\
0.920189922 & 0.930182952 & 0.92869332 & 0.921773082 & 0.933871572 & 0.92365675 & 0.931857587 & 0.9130443 & 0.928150977 & 0.933502435 & 0.94039889 & 0.952980478 & 0.908826678 & 0.917268224 & 0 \\
0.938431598 & 0.948651337 & 0.947074841 & 0.939889383 & 0.954166179 & 0.942680592 & 0.949985275 & 0.952079028 & 0.946482777 & 0.941901025 & 0.957134388 & 0.951845749 & 0.967300592 & 0.953951957 & 0.921171843 & 0 \\
0.929943012 & 0.940067654 & 0.938510739 & 0.93140216 & 0.945369935 & 0.934087334 & 0.941424242 & 0.935901263 & 0.940552802 & 0.934319651 & 0.945096604 & 0.943479142 & 0.968187769 & 0.910882858 & 0.968187769 & 0 \\
0.893712525 & 0.903423388 & 0.901965889 & 0.895220825 & 0.907334321 & 0.897212702 & 0.904980132 & 0.890667715 & 0.901433522 & 0.904760365 & 0.912992525 & 0.921847677 & 0.890234953 & 0.894317539 & 0.952980478 & 0.906031228 & 0.892005669 & 0 \\
0.882341079 & 0.891933934 & 0.890483871 & 0.883799825 & 0.896135023 & 0.885934766 & 0.893398551 & 0.88338039 & 0.889951574 & 0.891294748 & 0.901011025 & 0.906255675 & 0.886741492 & 0.886513525 & 0.921847677 & 0.906255675 & 0.888255291 & 0.952980478 & 0 \\
0.885949235 & 0.895586932 & 0.894119687 & 0.887383182 & 0.900149243 & 0.889697632 & 0.89698416 & 0.891101356 & 0.893577556 & 0.892955967 & 0.904321262 & 0.906031228 & 0.898307478 & 0.893769471 & 0.906255675 & 0.921847677 & 0.899590051 & 0.921847677 & 0.952980478 & 0 \\
0.904577082 & 0.91442297 & 0.912913736 & 0.90601071 & 0.919421583 & 0.908543109 & 0.915776799 & 0.913916402 & 0.912351756 & 0.90976248 & 0.922960017 & 0.921171843 & 0.925061423 & 0.916165998 & 0.906031228 & 0.952980478 & 0.926135893 & 0.906255675 & 0.921847677 & 0.952980478 & 0
\end{bmatrix} \quad (14)$$

for arXiv

$$C(G)=\begin{bmatrix} 0 \\ 10906 & 0 \\ 10906 & 12238 & 0 \\ 7994 & 11426 & 9584 & 0 \\ 9116 & 12446 & 11522 & 11998 & 0 \\ 6103 & 8635 & 7408 & 11516 & 9581 & 0 \\ 6170 & 8576 & 7646 & 9886 & 10402 & 10832 & 0 \\ 5789 & 8213 & 7004 & 11176 & 8983 & 11348 & 9466 & 0 \\ 4598 & 6455 & 5633 & 8101 & 7456 & 8531 & 9982 & 9982 & 0 \end{bmatrix}, R(G)=\begin{bmatrix} 0 \\ 0.989009671 & 0 \\ 0.989009671 & 0.997909305 & 0 \\ 0.986776508 & 0.997564283 & 0.995834589 & 0 \\ 0.987915686 & 0.998694132 & 0.997006015 & 0.99858543 & 0 \\ 0.984612654 & 0.995374935 & 0.993652718 & 0.9975901 & 0.996417877 & 0 \\ 0.983985507 & 0.994737786 & 0.993022971 & 0.99658302 & 0.995821163 & 0.996699517 & 0 \\ 0.983805931 & 0.994559533 & 0.99283845 & 0.996789054 & 0.995599819 & 0.996869637 & 0.995710459 & 0 \\ 0.974145475 & 0.984791859 & 0.983090931 & 0.986809365 & 0.985843175 & 0.98690696 & 0.987968112 & 0.987968112 & 0 \end{bmatrix} \quad (15)$$

$$C(G)=\begin{bmatrix} 0 \\ 1630029 & 0 \\ 1703615 & 1797573 & 0 \\ 1630029 & 1613608 & 1797573 & 0 \\ 1542636 & 1618453 & 1807545 & 1750457 & 0 \\ 1542636 & 1750457 & 1807545 & 1618453 & 1650886 & 0 \\ 1593905 & 1730826 & 1879637 & 1730826 & 1807545 & 1807545 & 0 \\ 1493103 & 1708373 & 1730826 & 1576693 & 1618453 & 1750457 & 1797573 & 0 \\ 1493103 & 1576693 & 1730826 & 1708373 & 1750457 & 1618453 & 1797573 & 1613608 & 0 \\ 1364732 & 1493103 & 1593905 & 1493103 & 1542636 & 1542636 & 1703615 & 1630029 & 1630029 & 0 \end{bmatrix},$$

$$R(G)=\begin{bmatrix} 0 \\ 0.998888414 & 0 \\ 0.998980326 & 0.999886271 & 0 \\ 0.998888414 & 0.999774956 & 0.999886271 & 0 \\ 0.998877926 & 0.999783728 & 0.999895033 & 0.999788564 & 0 \\ 0.998877926 & 0.999788564 & 0.999895033 & 0.999783728 & 0.999792877 & 0 \\ 0.998978762 & 0.99988545 & 0.999996298 & 0.99988545 & 0.999895033 & 0.999895033 & 0 \\ 0.998868407 & 0.999779196 & 0.99988545 & 0.999774384 & 0.999783728 & 0.999788564 & 0.999886271 & 0 \\ 0.998868407 & 0.999774384 & 0.99988545 & 0.999779196 & 0.999788564 & 0.999783728 & 0.999886271 & 0.999774956 & 0 \\ 0.997962328 & 0.998868407 & 0.998978762 & 0.998868407 & 0.998877926 & 0.998877926 & 0.998980326 & 0.998888414 & 0.998888414 & 0 \end{bmatrix} \quad (16)$$

$$C(G)=\begin{bmatrix} 0 \\ 97195903 & 0 \\ 95224345 & 97700975 & 0 \\ 68471291 & 85243051 & 72389836 & 0 \\ 81765418 & 87956993 & 91528781 & 85243051 & 0 \\ 90266101 & 85305365 & 80376470 & 66803146 & 89357441 & 0 \\ 75888313 & 90141767 & 72962504 & 62664034 & 73616663 & 87006599 & 0 \\ 55630271 & 64088005 & 59834158 & 54550292 & 78425719 & 58906828 & 57216160 & 0 \\ 46646959 & 55853312 & 45240473 & 39027886 & 46157399 & 52570571 & 78619511 & 39111712 & 0 \\ 36805526 & 44181082 & 38909740 & 35111372 & 47591158 & 37753048 & 38147704 & 73985111 & 26293558 & 0 \\ 55103884 & 66570359 & 54347075 & 47294935 & 57075608 & 60447596 & 84817535 & 56616889 & 64738385 & 37565551 & 0 \\ 44419573 & 53935106 & 43752029 & 38037175 & 45502883 & 48488765 & 68086373 & 43905379 & 78619511 & 29843929 & 82613609 & 0 \\ 38673818 & 49155670 & 39829060 & 35366156 & 43271482 & 37777984 & 40553464 & 54183173 & 28289782 & 73985111 & 39656041 & 32603839 & 0 \\ 55824986 & 68712598 & 57087172 & 50585744 & 63532858 & 57519532 & 67544968 & 80361791 & 50084656 & 52431338 & 82397047 & 62035153 & 53919878 & 0 \\ 38354561 & 47732011 & 38432506 & 33676628 & 40298761 & 39830236 & 50906680 & 42769022 & 49324360 & 30356297 & 59309605 & 76447999 & 35069339 & 63029153 & 0 \\ 36548575 & 45187517 & 37075322 & 32709184 & 40325579 & 37777766 & 45947168 & 47923855 & 38538872 & 32188381 & 55006964 & 53730416 & 34563511 & 76933771 & 76933771 & 0 \\ 62169293 & 81499063 & 63051778 & 55442036 & 63306853 & 58994104 & 65636320 & 64910945 & 46098496 & 54183173 & 63933604 & 53565064 & 73985111 & 85571681 & 59264669 & 56236810 & 0 \\ 39142009 & 50351699 & 39519290 & 34704580 & 40325729 & 38450564 & 45316568 & 41902144 & 37087952 & 32952049 & 47892578 & 50593346 & 42602563 & 57805345 & 75678943 & 51807776 & 75678943 & 0 \end{bmatrix},$$

for arXiv

$$R(G)=\begin{bmatrix}
0 \\
0.998759167 & 0 \\
0.997888637 & 0.998831068 & 0 \\
0.988649761 & 0.989829016 & 0.988743476 & 0 \\
0.998497644 & 0.999639033 & 0.998615131 & 0.989829016 & 0 \\
0.997831417 & 0.998761768 & 0.997707458 & 0.988696128 & 0.998588784 & 0 \\
0.99823193 & 0.999418781 & 0.99829551 & 0.989298801 & 0.999149186 & 0.998369387 & 0 \\
0.995479104 & 0.996658789 & 0.995559704 & 0.986637055 & 0.996532218 & 0.99555295 & 0.996389506 & 0 \\
0.986921884 & 0.988097291 & 0.986986856 & 0.978093832 & 0.987834571 & 0.987038888 & 0.988451792 & 0.98531231 & 0 \\
0.977349679 & 0.978512219 & 0.977425674 & 0.96865135 & 0.978344365 & 0.977419633 & 0.978260751 & 0.979952437 & 0.967401146 & 0 \\
0.996910929 & 0.998100087 & 0.996978586 & 0.987997308 & 0.997838642 & 0.99701114 & 0.998247511 & 0.995497004 & 0.987428731 & 0.977408268 & 0 \\
0.995409595 & 0.996597125 & 0.995477258 & 0.986509538 & 0.996336034 & 0.99550852 & 0.99673156 & 0.994001041 & 0.988451792 & 0.975948268 & 0.996418822 & 0 \\
0.978061888 & 0.979229696 & 0.978134748 & 0.969339638 & 0.979017282 & 0.978129308 & 0.97899122 & 0.978788477 & 0.968139971 & 0.979952437 & 0.978162363 & 0.976710252 & 0 \\
0.997849026 & 0.999041991 & 0.997919956 & 0.988933477 & 0.99878799 & 0.997922081 & 0.998869156 & 0.996832494 & 0.98785338 & 0.978708033 & 0.998183809 & 0.99665441 & 0.979451385 & 0 \\
0.99379685 & 0.99498471 & 0.993866613 & 0.984915239 & 0.994727677 & 0.993875606 & 0.994886697 & 0.992608773 & 0.984613224 & 0.974625014 & 0.994262182 & 0.994094091 & 0.975430777 & 0.995333862 & 0 \\
0.985963641 & 0.98714227 & 0.986033293 & 0.977153162 & 0.986889281 & 0.986038782 & 0.987008099 & 0.984872282 & 0.976467646 & 0.966996782 & 0.986359483 & 0.985518029 & 0.967763637 & 0.987789724 & 0.987789724 & 0 \\
0.997623648 & 0.998819192 & 0.997694807 & 0.988709566 & 0.998558447 & 0.99768986 & 0.998589028 & 0.996495189 & 0.987536994 & 0.978788477 & 0.997767668 & 0.99629546 & 0.979952437 & 0.999070811 & 0.995035014 & 0.987181369 & 0 \\
0.985858314 & 0.987038228 & 0.98592808 & 0.977048506 & 0.986781899 & 0.985930065 & 0.986875663 & 0.984711503 & 0.976317176 & 0.967042434 & 0.986159698 & 0.985346585 & 0.968017709 & 0.987335669 & 0.987648164 & 0.977711296 & 0.987648164 & 0
\end{bmatrix} \quad (17)$$

$$C(G)=\begin{bmatrix}
0 \\
3149069 & 0 \\
3043027 & 3228245 & 0 \\
2858734 & 3321779 & 3112111 & 0 \\
2914171 & 2906105 & 2641000 & 2984239 & 0 \\
2506019 & 2987917 & 2463944 & 2681189 & 2843153 & 0 \\
2177255 & 2653765 & 2289866 & 2783081 & 2228390 & 2326876 & 0 \\
1978064 & 2393149 & 1979801 & 2199512 & 2164634 & 2952505 & 2236627 & 0 \\
1905871 & 2302280 & 1901701 & 2102821 & 2096503 & 2899505 & 2076155 & 2937515 & 0 \\
2074988 & 2560606 & 2131142 & 2439176 & 2148470 & 2530210 & 2971729 & 2789797 & 2492735 & 0 \\
1717739 & 2112889 & 1746902 & 1966949 & 1807382 & 2251960 & 2211610 & 2346985 & 2657297 & 2814799 & 0 \\
2271161 & 2875501 & 2344652 & 2628131 & 2252564 & 2463286 & 3047311 & 2366140 & 2233490 & 3099751 & 2466163 & 0 \\
1533905 & 1918909 & 1573568 & 1767095 & 1560554 & 1811758 & 2023216 & 1810606 & 1879940 & 2273695 & 2569681 & 2569681 & 0
\end{bmatrix},$$

$$R(G)=\begin{bmatrix}
0 \\
0.998785534 & 0 \\
0.997904872 & 0.998868351 & 0 \\
0.998665745 & 0.999832626 & 0.998769443 & 0 \\
0.997866735 & 0.998822164 & 0.997755584 & 0.99874497 & 0 \\
0.998396555 & 0.99958142 & 0.998473139 & 0.999460195 & 0.998522703 & 0 \\
0.997407746 & 0.998594275 & 0.997491068 & 0.998503026 & 0.997486649 & 0.998440251 & 0 \\
0.997262475 & 0.998448091 & 0.997341081 & 0.998329235 & 0.997368123 & 0.998628207 & 0.997531008 & 0 \\
0.997100971 & 0.998286369 & 0.997179534 & 0.998167477 & 0.997206887 & 0.998469827 & 0.997364605 & 0.997789336 & 0 \\
0.998232714 & 0.999421856 & 0.998313816 & 0.999305625 & 0.998315246 & 0.999330267 & 0.998773854 & 0.998499791 & 0.998314301 & 0 \\
0.996262818 & 0.997449304 & 0.996343391 & 0.997332652 & 0.996348564 & 0.997397858 & 0.996745502 & 0.996573708 & 0.996773452 & 0.997747341 & 0 \\
0.99819973 & 0.999389659 & 0.998281204 & 0.999272656 & 0.998277917 & 0.999251244 & 0.998746631 & 0.998356165 & 0.998192727 & 0.999600658 & 0.99760711 & 0 \\
0.987309748 & 0.988486136 & 0.987389967 & 0.98837047 & 0.987390889 & 0.988392101 & 0.987819442 & 0.987540983 & 0.987558647 & 0.988738135 & 0.988866526 & 0.988866526 & 0
\end{bmatrix} \quad (18)$$

for arXiv

$$C(G)=\begin{bmatrix}
0 \\
633546984 & 0 \\
633546984 & 461787600 & 0 \\
633546984 & 461787600 & 461787600 & 0 \\
461787600 & 633546984 & 461787600 & 365166144 & 0 \\
461787600 & 633546984 & 461787600 & 365166144 & 633546984 & 0 \\
461787600 & 633546984 & 365166144 & 461787600 & 461787600 & 365166144 & 0 \\
461787600 & 633546984 & 365166144 & 461787600 & 365166144 & 329038344 & 633546984 & 0 \\
461787600 & 365166144 & 461787600 & 633546984 & 329038344 & 365166144 & 365166144 & 461787600 & 0 \\
461787600 & 365166144 & 633546984 & 461787600 & 365166144 & 461787600 & 329038344 & 365166144 & 633546984 & 0 \\
365166144 & 461787600 & 365166144 & 329038344 & 633546984 & 461787600 & 461787600 & 365166144 & 311658948 & 329038344 & 0 \\
365166144 & 461787600 & 365166144 & 329038344 & 461787600 & 365166144 & 365166144 & 461787600 & 329038344 & 311658948 & 633546984 & 0 \\
329038344 & 365166144 & 311658948 & 365166144 & 365166144 & 329038344 & 461787600 & 461787600 & 365166144 & 329038344 & 461787600 & 633546984 & 0 \\
365166144 & 365166144 & 329038344 & 461787600 & 329038344 & 311658948 & 461787600 & 633546984 & 461787600 & 365166144 & 461787600 & 633546984 & 0 \\
365166144 & 365166144 & 329038344 & 365166144 & 461787600 & 311658948 & 329038344 & 365166144 & 461787600 & 633546984 & 461787600 & 461787600 & 633546984 & 0 \\
329038344 & 311658948 & 365166144 & 365166144 & 329038344 & 365166144 & 329038344 & 365166144 & 461787600 & 461787600 & 365166144 & 365166144 & 461787600 & 461787600 & 633546984 & 0 \\
365166144 & 329038344 & 461787600 & 365166144 & 365166144 & 461787600 & 311658948 & 329038344 & 461787600 & 633546984 & 365166144 & 329038344 & 365166144 & 365166144 & 461787600 & 633546984 & 0 \\
365166144 & 365166144 & 461787600 & 329038344 & 461787600 & 633546984 & 329038344 & 365166144 & 311658948 & 365166144 & 365166144 & 329038344 & 329038344 & 365166144 & 633546984 & 461787600 & 633546984 & 0 \\
329038344 & 365166144 & 365166144 & 311658948 & 461787600 & 461787600 & 365166144 & 329038344 & 329038344 & 365166144 & 633546984 & 461787600 & 461787600 & 365166144 & 365166144 & 461787600 & 461787600 & 633546984 & 0 \\
311658948 & 329038344 & 329038344 & 329038344 & 365166144 & 365166144 & 365166144 & 365166144 & 365166144 & 365166144 & 461787600 & 461787600 & 633546984 & 461787600 & 461787600 & 633546984 & 461787600 & 461787600 & 633546984 & 0
\end{bmatrix}$$

$$R(G)=\begin{bmatrix}
0 \\
0.997505449 & 0 \\
0.997505449 & 0.997211982 & 0 \\
0.997505449 & 0.997211982 & 0.997211982 & 0 \\
0.997211982 & 0.997505449 & 0.997211982 & 0.997142738 & 0 \\
0.997211982 & 0.997505449 & 0.997211982 & 0.997142738 & 0.997505449 & 0 \\
0.997211982 & 0.997505449 & 0.997142738 & 0.997211982 & 0.997211982 & 0.997142738 & 0 \\
0.997211982 & 0.997211982 & 0.997142738 & 0.997505449 & 0.997142738 & 0.997127062 & 0.997505449 & 0 \\
0.997211982 & 0.997142738 & 0.997211982 & 0.997505449 & 0.997127062 & 0.997142738 & 0.997142738 & 0.997211982 & 0 \\
0.997211982 & 0.997142738 & 0.997505449 & 0.997211982 & 0.997142738 & 0.997211982 & 0.997127062 & 0.997142738 & 0.997505449 & 0 \\
0.997142738 & 0.997211982 & 0.997142738 & 0.997127062 & 0.997505449 & 0.997211982 & 0.997211982 & 0.997142738 & 0.997120399 & 0.997127062 & 0 \\
0.997142738 & 0.997211982 & 0.997142738 & 0.997127062 & 0.997211982 & 0.997142738 & 0.997142738 & 0.997211982 & 0.997127062 & 0.997120399 & 0.997505449 & 0 \\
0.997127062 & 0.997142737 & 0.997120399 & 0.997142737 & 0.997142737 & 0.997127062 & 0.997211982 & 0.997211982 & 0.997142737 & 0.997127062 & 0.997211982 & 0.997505449 & 0 \\
0.997142738 & 0.997142738 & 0.997127062 & 0.997211982 & 0.997127062 & 0.997120399 & 0.997211982 & 0.997505449 & 0.997211982 & 0.997142737 & 0.997142737 & 0.997211982 & 0.997505449 & 0 \\
0.997142738 & 0.997127062 & 0.997142738 & 0.997142738 & 0.997211982 & 0.997120399 & 0.997127062 & 0.997142738 & 0.997505449 & 0.997211982 & 0.997142737 & 0.997142737 & 0.997211982 & 0.997505449 & 0 \\
0.997127062 & 0.997120399 & 0.997142737 & 0.997142737 & 0.997127062 & 0.997142737 & 0.997127062 & 0.997142737 & 0.997211982 & 0.997211982 & 0.997142737 & 0.997142737 & 0.997211982 & 0.997211982 & 0.997505449 & 0 \\
0.997142738 & 0.997127062 & 0.997211982 & 0.997142738 & 0.997142738 & 0.997211982 & 0.997120399 & 0.997127062 & 0.997211982 & 0.997505449 & 0.997142737 & 0.997127062 & 0.997142737 & 0.997142737 & 0.997211982 & 0.997505449 & 0 \\
0.997142738 & 0.997142738 & 0.997211982 & 0.997127062 & 0.997211982 & 0.997505449 & 0.997127062 & 0.997142738 & 0.997120399 & 0.997142738 & 0.997142737 & 0.997127062 & 0.997127062 & 0.997142737 & 0.997505449 & 0.997211982 & 0.997505449 & 0 \\
0.997127062 & 0.997142738 & 0.997142737 & 0.997120399 & 0.997211982 & 0.997211982 & 0.997142737 & 0.997127062 & 0.997127062 & 0.997142737 & 0.997505449 & 0.997211982 & 0.997211982 & 0.997142737 & 0.997142737 & 0.997211982 & 0.997211982 & 0.997505449 & 0 \\
0.997120399 & 0.997127062 & 0.997127062 & 0.997127062 & 0.997142738 & 0.997142738 & 0.997142737 & 0.997142737 & 0.997142737 & 0.997142737 & 0.997211982 & 0.997211982 & 0.997505449 & 0.997211982 & 0.997211982 & 0.997505449 & 0.997211982 & 0.997211982 & 0.997505449 & 0
\end{bmatrix}$$

(19)

$$C(G)=\begin{bmatrix}
0 \\
9857088 & 0 \\
6487083 & 10087938 & 0 \\
4357917 & 6487083 & 9857088 & 0 \\
8386098 & 10847544 & 8963223 & 6501168 & 0 \\
6501168 & 8963223 & 10847544 & 8386098 & 11140740 & 0 \\
4781115 & 6669711 & 8599744 & 9857088 & 7859463 & 10847544 & 0 \\
6487083 & 6669711 & 5723992 & 4308939 & 8963223 & 7219257 & 5429359 & 0 \\
6501168 & 7859463 & 7219257 & 5578152 & 11140740 & 9479332 & 7219257 & 10847544 & 0 \\
5578152 & 7219257 & 7859463 & 6501168 & 9479332 & 11140740 & 8963223 & 7859463 & 11140740 & 0 \\
4308939 & 5723992 & 6669711 & 6487083 & 7219257 & 8963223 & 10087938 & 5702382 & 7859463 & 10847544 & 0 \\
4357917 & 4781115 & 4308939 & 3329245 & 6501168 & 5578152 & 4308939 & 9857088 & 8386098 & 6501168 & 4781115 & 0 \\
4781115 & 5702382 & 5429359 & 4308939 & 7859463 & 7219257 & 5723992 & 8599744 & 10847544 & 8963223 & 6669711 & 9857088 & 0 \\
4308939 & 5429359 & 5702382 & 4781115 & 7219257 & 7859463 & 6669711 & 6669711 & 8963223 & 10847544 & 8599744 & 6487083 & 10087938 & 0 \\
3329245 & 4308939 & 4781115 & 4357917 & 5578152 & 6501168 & 6487083 & 4781115 & 6501168 & 8386098 & 9857088 & 4357917 & 6487083 & 9857088 & 0 \\
9857088 & 8599744 & 6669711 & 4781115 & 10847544 & 7859463 & 5702382 & 10087938 & 8963223 & 7219257 & 5429359 & 6487083 & 6669711 & 5723992 & 4308939 & 0
\end{bmatrix},$$

for arXiv

$$R(G)=\begin{bmatrix}
0 \\
0.987831149 & 0 \\
0.985405124 & 0.995666375 & 0 \\
0.975419048 & 0.985405124 & 0.987831149 & 0 \\
0.987604335 & 0.997482238 & 0.997093665 & 0.98717353 & 0 \\
0.98717353 & 0.997093665 & 0.997482238 & 0.987604335 & 0.999277779 & 0 \\
0.984946496 & 0.994857083 & 0.995421314 & 0.987831149 & 0.997000825 & 0.997482238 & 0 \\
0.985405124 & 0.994857083 & 0.994657741 & 0.984798819 & 0.997093665 & 0.996868109 & 0.994639901 & 0 \\
0.98717353 & 0.997000825 & 0.996868109 & 0.98699616 & 0.999277779 & 0.999095757 & 0.996868109 & 0.997482238 & 0 \\
0.98699616 & 0.996868109 & 0.997000825 & 0.98717353 & 0.999095757 & 0.999277779 & 0.997093665 & 0.997000825 & 0.999277779 & 0 \\
0.984798819 & 0.994657741 & 0.994857083 & 0.985405124 & 0.996868109 & 0.997093665 & 0.995666375 & 0.994738987 & 0.997000825 & 0.997482238 & 0 \\
0.975419048 & 0.984946496 & 0.984798819 & 0.97504635 & 0.98717353 & 0.98699616 & 0.984798819 & 0.987831149 & 0.987604335 & 0.98717353 & 0.984946496 & 0 \\
0.984946496 & 0.994738987 & 0.994639901 & 0.984798819 & 0.997000825 & 0.996868109 & 0.994657741 & 0.995421314 & 0.997482238 & 0.997093665 & 0.994857083 & 0.987831149 & 0 \\
0.984798819 & 0.994639901 & 0.994738987 & 0.984946496 & 0.996868109 & 0.997000825 & 0.994857083 & 0.994857083 & 0.997093665 & 0.997482238 & 0.995421314 & 0.985405124 & 0.995666375 & 0 \\
0.97504635 & 0.984798819 & 0.984946496 & 0.975419048 & 0.98699616 & 0.98717353 & 0.985405124 & 0.984946496 & 0.98717353 & 0.987604335 & 0.987831149 & 0.975419048 & 0.985405124 & 0.987831149 & 0 \\
0.987831149 & 0.995421314 & 0.994857083 & 0.984946496 & 0.997482238 & 0.997000825 & 0.994738987 & 0.995666375 & 0.997093665 & 0.996868109 & 0.994639901 & 0.985405124 & 0.994857083 & 0.994657741 & 0.984798819 & 0
\end{bmatrix} \quad (20)$$

for arXiv

for arXiv